\documentclass[12pt]{article}
\usepackage{amssymb}
\usepackage{amsmath}
\usepackage{epsf}
\usepackage{epsfig}
\usepackage{psfrag}
\usepackage{rotating}
\sffamily
\allowdisplaybreaks[1]

\newcommand{\lsim}{
\mathrel{\hbox{\rlap{\hbox{\lower4pt\hbox{$\sim$}}}\hbox{$<$}}}}

\newcommand{\gsim}{
\mathrel{\hbox{\rlap{\hbox{\lower4pt\hbox{$\sim$}}}\hbox{$>$}}}}

\newcommand{\vcb}{|V_{cb}|}
\newcommand{\vtd}{|V_{td}|}

\newcommand{\gev}{\, {\rm GeV}}

\newcommand{\tev}{\, {\rm TeV}}

\newcommand{\mtb}{\overline{m}_{\rm t}}

\newcommand{\mw}{M_{\rm W}}

\newcommand{\be}{\begin{equation}}
\newcommand{\ee}{\end{equation}}
\newcommand{\bi}{\begin{itemize}}
\newcommand{\ei}{\end{itemize}}
\newcommand{\ord}{{\cal O}}

\def\kpn{K^+\rightarrow\pi^+\nu\bar\nu}
\def\klpn{K_{\rm L}\rightarrow\pi^0\nu\bar\nu}

\begin{document}
\begin{titlepage}
\vspace*{-0.5truecm}

\begin{flushright}
TUM-HEP-636/06\\
FERMILAB-PUB-06-172-T
\end{flushright}

\vspace*{0.3truecm}

\begin{center}
\boldmath

{\Large{\bf Rare $K$ and $B$ Decays  

\vspace{0.3truecm}

in the Littlest Higgs Model without T-Parity}}

\unboldmath
\end{center}

\vspace{0.4truecm}

\begin{center}
{\large \bf Andrzej J. Buras${}^a$, Anton Poschenrieder${}^a$, \\ 
\vspace{0.3truecm}
Selma Uhlig${}^a$ and William A. Bardeen${}^b$}
\vspace{0.4truecm}

${}^a$ {\sl Physik Department, Technische Universit\"at M\"unchen,
D-85748 Garching, Germany}

\vspace{0.2truecm}

${}^b$ {\sl Theoretical Physics Department, Fermilab, Batavia, IL 60510, 
USA}

\vspace{0.2truecm}

\end{center}

\vspace{0.6cm}
\begin{abstract}
\vspace{0.2cm}\noindent
We analyze rare $K$ and $B$ decays in the Littlest Higgs (LH) model
{\it without} T-parity. We find that the final result for the
$Z^0$-penguin contribution contains a divergence that is generated
by the one-loop radiative corrections to the currents corresponding
to the dynamically broken generators. Including an estimate of these
logarithmically enhanced terms, we calculate the branching ratios
for the decays $\kpn$, $\klpn$,
$B_{s,d}\to\mu^+\mu^-$ and $B\to X_{s,d}\nu\bar\nu$. We find that for the high
energy scale $f=\mathcal{O}\left(2-3\right) \textrm{TeV}$, as required by the
electroweak precision studies, the enhancement of all branching ratios
amounts to at most $15 \%$ over the SM values. On the technical side we
identify a number of errors in the existing Feynman rules in the LH model
without T-parity that could have some impact on other analyses
present in the literature. Calculating penguin and box diagrams in the
unitary gauge, we find divergences in both contributions that are cancelled
in the sum except for the divergence mentioned above.

\end{abstract}

%
%
%
\end{titlepage}
\thispagestyle{empty}
\vbox{}

\setcounter{page}{1}
\pagenumbering{roman}



\setcounter{page}{1}
\pagenumbering{arabic}
\section{Introduction}\label{sec:intro}
The Little Higgs models \cite{LH1}-\cite{LH5} offer
an attractive and a rather simple solution to  the gauge hierarchy problem.
In these models the electroweak Higgs boson is regarded as a
pseudo-Goldstone boson of a certain global
symmetry that is broken spontaneously at a scale
$\Lambda \sim 4 \pi f \sim \mathcal{O}\left(\textrm{10 TeV}\right)$,
 much higher than the vacuum expectation value $v$ of the standard Higgs
doublet. The Higgs field remains then light,
being protected by
the approximate global symmetry from acquiring quadratically divergent
contributions to its mass at the one-loop level.
On the diagrammatic level the new heavy particles present in these models
cancel, analogously to supersymmetric
particles, the quadratic divergencies in question.
Reviews on the Little Higgs models can be found in \cite{LHREV}.

One of the simplest models of this type is the ``Littlest Higgs'' model 
\cite{LH4} (LH) in which, in addition to the Standard Model (SM) particles, 
new charged heavy vector bosons ($W_H^\pm$), a neutral heavy vector 
boson ($Z^{0}_{H}$), a heavy photon ($A^{0}_{H}$), a heavy top quark ($T$) 
 and charged and neutral heavy Higgs scalars 
are present. Among the scalars only the single charged scalar ($\Phi^\pm$) 
is important in principle for rare decays. 
The details of this model including the Feynman rules have been 
worked out in \cite{Logan} and the constraints from various processes, 
in particular from electroweak precision observables and direct new particles
searches, have been extensively discussed in \cite{Logan}-\cite{PHEN6}. 
It has been found that 
except for the heavy photon $A^{0}_{H}$, that could still be as ``light'' as 
$500~\textrm{GeV}$, the masses of the remaining particles are constrained 
to be significantly larger than $1~ \textrm{TeV}$.

Much less is known about the flavour changing neutral current (FCNC)
processes in the LH model. As these processes played an essential role in the
construction of the SM and in the tests of its extensions, it is important to
check whether the LH model is consistent with the existing data on FCNC
processes and whether the deviations from the SM expectations predicted in
this model are sufficiently large so that they could be detected in present
and future experiments.

In \cite{BPU04} we have calculated
the $K^{0}-\bar{K}^{0}$, $B_{d,s}^{0}-\bar{B}_{d,s}^{0}$ 
mixing mass differences $\Delta M_K$, $\Delta M_{d,s}$ and the 
CP-violating parameter $\varepsilon_{K}$ in the LH model. 
We have found that
even for $f/v$ as low as $5$, the enhancement of 
$\Delta M_{d}$ 
amounts to at most $20\%$ for the Yukawa parameter $x_{L} \leq 0.8$. Similar 
comments apply to $\Delta M_s$ and $\varepsilon_{K}$.
The correction to $\Delta M_{K}$ is negligible.
These results have been confirmed in \cite{IND1}. 
Larger effects could be present in $D^0-\bar D^0$ mixing \cite{QUAD}, where 
in contrast to  processes involving external down quarks, FCNC transitions
are already present at the tree level. But as analyzed in \cite{fajfer} these
effects are small.

On the other hand we have pointed out in \cite{BPU04, DEC05} that for
$0.80 \leq x_{L} \leq 0.95$ and $f/v \leq 10$, which is still allowed
by the electroweak precision studies, the non-decoupling effects of
the heavy $T$ can significantly suppress the CKM element $\vtd$ and 
the angle $\gamma$ in the unitarity triangle and 
simultaneously enhance $\Delta M_{s}$. The recent data from CDF and
D$\emptyset$ collaborations \cite{CDF,D0} disfavour this possibility, although
in view of large non-perturbative uncertainties in the evaluation of
$\Delta M_{s}$ nothing conclusive can be said at present \cite{BBGT, FB}.

Concerning FCNC decay processes only $B\to X_s\gamma$ and $\klpn$ have been
considered so far in the literature. 
While in \cite{BSG} the LH corrections to the the decay $B\to X_s\gamma$ have 
 been found to be small, a large enhancement of the
branching ratio for $\klpn$ relative to the SM expectations has been found in
\cite{IND2}.

In the present paper we extend our study of FCNC processes in the LH model to
the rare decays $\kpn$, $\klpn$, $B_{s,d}\to\mu^+\mu^-$ and $B\to
X_{s,d}\nu\bar\nu$. We also briefly discuss the decay 
$B\to X_s\gamma$. 

The analysis of the rare decays in question turned out to be much more
involved than the one of particle--antiparticle mixing due to the presence of
many more diagrams, in particular the $Z^0$, $Z^{0}_{H}$ and $A^{0}_{H}$ 
penguins that were 
absent in our previous study. In order to reduce the number of contributing
diagrams we have performed all calculations in the unitary 
 gauge for the $W_L^{\pm}$ and $W_H^{\pm}$ propagators which
has the nice virtue that only exchanges of physical particles have to be
considered. 

Already in \cite{BPU04,DEC05} we have found that the box diagrams 
contributing to
particle-antiparticle mixing were divergent in the unitary gauge but these
divergences cancelled each other after
the unitarity of the CKM matrix
in the SM has been used and the contribution of the heavy $T$  included at
$\ord(v^2/f^2)$. Simply, the GIM mechanism \cite{GIM} was sufficiently 
effective 
to remove these divergences. In the case of rare decays, to which also
penguin diagrams contribute, the cancellation of divergences, even in the SM,
is more involved due to a different structure of the diagrams. It turns out
that the contributions of box diagrams to decay amplitudes remain divergent
even after the GIM mechanism has been used. However the full contribution of 
penguin diagrams is also divergent and in the SM this divergence cancels the
one from the box diagrams.

On the other hand in our analysis of the complete set of contributions to
the weak decay amplitudes in the LH model we find a remaining divergence
in our unitary gauge calculation.
While all divergent contributions from the box diagrams are exactly 
cancelled by gauge related divergences of the vertex contributions as in 
the SM,
there is a remaining divergence generated by the radiative corrections 
to the quark vertex.   The origin of this divergence can be traced to 
the structure of charge renormalization for the currents associated 
with the dynamically broken generators.  
For linearly realized symmetries, 
current conservation implies that the charges are 
not renormalized by radiative corrections.  
However, conserved currents associated with dynamically broken charges 
are not protected from renormalization and the charge vertex 
can be modified by the radiative corrections.   
The currents remain conserved because there is a 
corresponding modification of the Goldstone boson contribution 
to the current matrix element.   
In the nonlinear sigma model used to describe the little Higgs theory, 
these contributions can be divergent and depend on the UV completion of 
the theory. In a linear sigma model, the UV cutoff would be identified 
with symmetry breaking within the meson multiplet and related 
to the masses of the heavy partners to the Goldstone bosons.
A more general UV completion may even include charge renormalization 
at tree level.

This mechanism is analogous to the dynamics associated
with renormalization of axial-vector charge, $G_A$,
in the constituent quark model.
In particular, Peris  \cite{Peris}
has shown that the axial charge of the constituent quark is suppressed
by the one-loop radiative corrections in agreement with the quark model 
description of the axial charge of the physical baryons.  
The divergent contributions to the weak decay amplitudes will be discussed 
in more detail in Section \ref{sec:Bardeen}.

In the process of our analysis
we found several errors in certain  Feynman rules for the $v^2/f^2$ 
corrections to the $Z^0\bar f f$ vertices and in the 
vertices involving the heavy $T$ that were given
in \cite{Logan}. Without correcting these rules the final result would 
contain many more divergences and parametric dependencies that 
should be absent.

We are not the first to consider the decay $\klpn$  within
the LH model. 
In \cite{IND2} this decay has been analyzed with the result that its
branching ratio could be enhanced by a factor of two or more by LH
contributions relative to the SM expectations \cite{BSU}. 
This would be a very nice result as an enhancement of this
size in a theoretically clean decay $\klpn$ could clearly be distinguished
from the SM in future experiments.

Unfortunately our analysis of $\klpn$ presented here does not
confirm the findings of \cite{IND2}. This possibly can be traced back to
the fact that
 these authors used the Feynman rules of \cite{Logan} that according to
our analysis cannot give correct results for rare decay branching ratios 
in question. 
There is also the following qualitative difference between the
final results presented in \cite{IND2} and ours. It is related to the 
additional weak mixing angle $s^\prime$, present in the LH model, that is
analogous to $\sin\theta_w$ in the SM. The short distance function $X$
relevant for FCNC processes with $\nu\bar\nu$ in the final state cannot
depend on  $\sin\theta_w$ and $s^\prime$ due to current conservation. Our
results for $X$ and the function $Y$, relevant for the FCNC processes with 
$l^+l^-$  in the final state, are indeed independent 
of $\sin\theta_w$ and $s^\prime$, while the numerical results presented in
\cite{IND2} show a clear $s^\prime$ dependence.

The main goal of our paper is the calculation of the LH contributions to the
short distance functions $X$ and $Y$ \cite{IL,PBE}. This will allow us to 
compute the impact of these contributions on various rare $K$ and $B$ decay 
branching ratios, which enter universally all decays in models with minimal 
flavour violation \cite{MFV} such as the LH model considered here. Our main 
findings for $\kpn$, $\klpn$, $B_{s} \to \mu^{+} \mu^{-}$ in the limit 
$x_{L} \approx 1$ have been summarized in \cite{DEC05}. In this limit
$\Delta M_{s}$ can be significantly enhanced with respect to the SM,
although this limit seems to be disfavoured by the recent CDF and D$\emptyset$
data. Here we present the details of these investigations in the full
space of the parameters involved, that requires the inclusion of many
more diagrams. We also extend our analysis to other rare decays and to 
the $B \to X_{s} \gamma$ decay.

Our paper is organized as follows. In 
Section~\ref{sec:LH} we recall briefly those elements of the LH model
that are necessary for the discussion of our calculation. 
In particular we give the $U(1)$ charges for quarks and leptons and 
present the list of the relevant Feynman rules that at various places differ
from those found in \cite{Logan}. 

In Section~\ref{sec:XYSM} we discuss the functions $X$ and $Y$ within the 
SM, presenting for the first 
 time  the expressions for the $Z^0$ penguin 
function $C$ and the relevant box functions $B^{\nu\bar\nu}$ and 
$B^{\mu\bar\mu}$ in the unitary gauge. All these functions are divergent 
but inserting them in the expressions for $X$ and $Y$ one recovers the known 
finite results. The result for $C$ in the unitary gauge will be particularly
relevant for our LH calculation. 

Section~\ref{sec:XYLH} is devoted to the calculation 
of $X$ and $Y$ within the LH model. We group the diagrams in six
classes. We present analytic results for each class and for the full
correction to the functions $X$ and $Y$. In three of these classes the 
divergences in diagrams belonging to a given class cancel each other. In
the remaining classes the divergences with a simple structure remain.
In Section \ref{sec:Bardeen} we discuss in more detail the origin of the
leftover divergences that could be of interest for other little Higgs models.
We also give an estimate of these logarithmic enhanced terms, which
turn out to be small.

In Section \ref{sec:Pheno} we present the numerical results for various 
branching ratios. Thanks to  compendia  in \cite{BBL,BSW,BPSW}, that 
give various branching ratios in terms of the functions $X$ and $Y$ we 
do not have to list once again all these formulae so that this 
section can be kept brief in spite of many decays involved. 
 
In Section \ref{sec:Bsg} we first briefly discuss the $B\to X_s\gamma$ decay 
confirming basically the result of \cite{BSG}. 
A brief summary of our paper is given in Section \ref{sec:Summ}.
Some technicalities are relegated to the appendices.

\section{Aspects of the Littlest Higgs Model}\label{sec:LH}
\setcounter{equation}{0}
\subsection{Gauge Boson Sector}
Let us first recall certain aspects of the LH model that are relevant for our
work. The full exposition can be found in the original paper \cite{LH4} and 
in \cite{Logan}. We will follow as far as possible the
notations of \cite{Logan}.

The global symmetry in the LH
model is a $SU(5)$ with a locally gauged subgroup
\be
G_1\otimes G_2=\left[SU(2)_1\otimes
  U(1)_1\right]\otimes\left[SU(2)_2\otimes U(1)_2\right]~.
\ee

In the process of the spontaneous breakdown of
the global $SU(5)$  at a scale $\Lambda$ $\sim
4\pi f \sim 10~\tev$ to a global $SO(5)$, the
gauge group $G_1\otimes G_2$ is broken down to the electroweak
SM gauge group $SU(2)_L\otimes U(1)_Y$. 
The resulting mass eigenstates in the gauge boson sector are
\begin{align}
W\;&=s\, W_1+c\, W_2, \qquad& B\;&=s'B_1+c' B_2,&\\
W'&=-c\, W_1+s\, W_2, \qquad& B'&=-c' B_1+s' B_2.&
\end{align}
Here $W_1$ and $W_2$ represent symbolically the three gauge bosons of 
$SU(2)_1$ and $SU(2)_2$, respectively. $B_1$ and $B_2$ are the corresponding
gauge bosons of $U(1)_1$ and $U(1)_2$. 

Note that $W=W_1$, $W'=W_2$,
$B=B_1$, $B'=B_2$ for $s=1$ and $s'=1$ and not for $s=0$ and $s'=0$. Thus in 
fact $s$, $s'$, $c$ and $c'$ are the sines and cosines of the mixing angles
plus $90$ degrees and not of the mixing angles as usually done in other
cases in the literature. The replacements $s\to c$ and $c\to s$ would 
be certainly a better choice. However, in order not to mix up the comparison 
of Feynman rules presented here with the ones of \cite{Logan} we will 
use the conventions of these authors, remembering that the mixing between
various groups is absent for $s=1$ and $s'=1$.

We have
\be
s=\frac{g_2}{\sqrt{g_1^2+g_2^2}}, \qquad
  c=\frac{g_1}{\sqrt{g_1^2+g_2^2}},
\ee
\be
  s'=\frac{g_2'}{\sqrt{g_1'^2+g_2'^2}},\qquad
  c'=\frac{g_1'}{\sqrt{g_1'^2+g_2'^2}},
\ee
where $g_{1, 2}$ are the $SU(2)_{1, 2}$ coupling constants and  
$g_{1,2}'$ the ones of the $U(1)_{1, 2}$.

The $W'$ and $B'$ gauge bosons receive the heavy masses 
\be\label{heavymasses}
m_{W'}=\frac{f}{2}\sqrt{g_1^2+g_2^2}=\frac{g}{2 s c} f, \qquad
m_{B'}=\frac{f}{2 \sqrt{5}}\sqrt{g_1'^2+g_2'^2}=\frac{g'}{2\sqrt{5} s'
  c'} f,
\ee
while the fields $W$ and $B$  remain massless at
this stage and can be identified as the SM gauge bosons with the
couplings $g$ and $g'$ given by
\be
g=g_1 s=g_2 c, \qquad g'=g_1' s'=g_2' c'.
\ee

In the second step of the gauge symmetry breaking the SM group is broken down
to $U(1)_Q$. The details of this breakdown are presented in \cite{Logan}.
As our results differ at certain places from those given in this paper, 
we give below the most relevant formulae summarizing subsequently the 
differences.

The mass eigenstates of the gauge bosons can be obtained by
diagonalizing 
\begin{align}\label{Lmasses}
{\mathcal L}_{masses}=&\;W_\mu^{'a}W^{'a \mu} \left(\frac{m_{W'}^2}{2}-\frac{1}{8}g^2
v^2\right)+(W_\mu^1 W^{1 \mu}+W_\mu^2 W^{2 \mu})\left(\frac{1}{8} g^2
v^2 \left(1+r\frac{v^2}{ f^2}\right)\right)\nonumber\\
&+W_\mu^3 W^{3 \mu}\left(\frac{1}{8} g^2
v^2 \left(1+r\frac{v^2}{f^2} \right)\right)+W_\mu^{a}W^{'a \mu}
\left(-\frac{1}{4}g^2 v^2 \frac{(c^2-s^2)}{2 s c}\right)\nonumber\\
&+B_\mu' B^{'
  \mu}\left(\frac{m_{B'}^2}{2}-\frac{1}{8}g'^2v^2\right)+B_\mu B^{
  \mu} \left(\frac{1}{8} g'^2
v^2 \left(1+r\frac{v^2}{f^2} \right)\right)\nonumber\\
&+B_\mu B^{' \mu}
\left(-\frac{1}{4}g'^2 v^2 \frac{(c'^2-s'^2)}{2 s' c'}\right)+W_\mu^3
B^{\mu}\left(\frac{1}{4}g g'v^2 \left(1+r\frac{v^2}{ f^2} \right)\right)
\nonumber\\
&+W_\mu^{' 3}
B^{'\mu}\left(-\frac{1}{8}g g'v^2 \left(\frac{cs'}{sc'}+\frac{sc'}{cs'}\right)\right)+W^3_\mu B^{' \mu}
\left(-\frac{1}{4}g g' v^2 \frac{(c'^2-s'^2)}{2 s'
    c'}\right)\nonumber\\
&+W_\mu^{' 3}B^{\mu}
\left(-\frac{1}{4}g g' v^2 \frac{(c^2-s^2)}{2 s c}\right),
\end{align}
with $v$  denoting the vacuum expectation of
the neutral components of the complex doublet. In our analysis we will set the 
vacuum expectation of  the Higgs triplet $v^{\prime}$ to zero.
For the parameter $r$ in  (\ref{Lmasses}) we find $r=-1/6$ that agrees with
(A30) in \cite{Logan} but differs from \cite{PHEN1}, where $r=1/2$  
can be found.
This difference 
has no direct impact on our calculation and as discussed in \cite{PHEN1} can
be absorbed through the redefinition of the parameters involved.

The final mass eigenstates of
the charged gauge bosons are $W_L^{\pm}$ and $W_H^{\pm}$ where the
indices $L$ and
$H$ stand for ``light'' and ``heavy''. The mass eigenstates are
\be
W_L=W+\frac{v^2}{2 f^2}sc(c^2-s^2) W',\qquad W_H=W'-\frac{v^2}{2
  f^2}sc(c^2-s^2) W,
\ee
and  the corresponding masses read ($r=-1/6$)
\begin{eqnarray}
M^2_{W_L^{\pm}}~&=&~m_w^2 \left(1-\frac{v^2}{f^2} \left(-r+\frac{1}{4}
(c^2-s^2)^2\right) \right)\label{massWL}\\
M^2_{W_H^{\pm}}~&=&~m_w^2 \left(\frac{f^2}{s^2 c^2 v^2}-1\right).\label{massWH}
\end{eqnarray}
The mass of the $W^{\pm}$ boson in the SM is given by $m_w\equiv
gv/2$. 

The
neutral gauge boson mass eigenstates are $A_L$, $Z_L$, $A_H$ and $Z_H$
given by
\begin{align}
A_L&=-s_w W^3+c_w B,\nonumber\\
Z_L&= c_w W^3+s_w
B+x_Z^{W'}\frac{v^2}{f^2}W^{'3}+x_Z^{B'}\frac{v^2}{f^2}B^{'},\nonumber\\
A_H&=B'+x_H \frac{v^2}{f^2}W^{'3}-x_Z^{B'}\frac{v^2}{f^2}(c_w W^3+s_w
B),\nonumber\\
Z_H&=W^{'3}-x_H \frac{v^2}{f^2}B'-x_Z^{W'}\frac{v^2}{f^2}(c_w W^3+s_w
B),\label{eigenvectors}
\end{align}
with
\begin{align}
x_H\;&=\frac{5}{2}g g'\frac{scs'c'(c^2 s^{'2}+s^2 c^{'2})}{(5 g^2
  s^{'2}c^{'2}-g^{'2}s^2c^2)}\,,\nonumber\\
x_Z^{W'}&=\frac{1}{2c_w}sc(c^2-s^2)\,,\nonumber\\
x_Z^{B'}&=\frac{5}{2s_w}s'c'(c^{'2}-s^{'2})\,.\label{xWxB}
\end{align}
Here
\be
s_w=\frac{g'}{\sqrt{g^2+g^{'2}}}\,,\qquad
c_w=\frac{g}{\sqrt{g^2+g^{'2}}}\,.
\ee
are the sine and the cosine of the Weinberg
angle describing the weak mixing in the SM.

$A_L$ and $Z_L$ are the SM photon and $Z^0$ boson and $A_H$ and
$Z_H$ the new heavy photon and heavy $Z^0$ boson, respectively. Their masses 
are given by ($r=-1/6$)
\begin{eqnarray}
M^2_{A_L}~&=&~0,\\
M^2_{Z_L}~&=&~m_z^2 \left(1-\frac{v^2}{f^2}\left(-r+\frac{1}{4}
(c^2-s^2)^2+\frac{5}{4} (c'^2-s'^2)^2\right) \right),
\label{massZL}\\
M^2_{A_H}~&=&~m_z^2 s_w^2 \left(\frac{f^2}{5s'^2 c'^2 v^2}-1\right),\label{MAH}\\
M_{Z_H}^2~&=&~m_w^2\left(\frac{f^2}{s^2 c^2 v^2}-1\right),\label{MZH}
\end{eqnarray}
where $m_z$ is the SM $Z^0$ boson mass with $m_z\equiv gv/(2
c_w)$. 

It is evident from (\ref{massWL}) and (\ref{massZL}) that
the tree level SM relation
\be\label{custodial}
\frac{m_w^2}{m_z^2}=c_w^2\,
\ee
is not valid for the $W_L^\pm$ and $Z_L^0$ masses.
To $\mathcal{O}(v^2/f^2)$ we have
 \cite{Logan}
\be\label{break}
\frac{M^2_{W_L^\pm}}{M^2_{Z_L}}=c_w^2 \left(1+ \frac{v^2}{f^2}
\frac{5}{4} (c'^2-s'^2)^2\right)
\ee
which manifests the breaking of the custodial $SU(2)$ in the LH model. 
Formula (\ref{break}) will 
play an important role in our analysis.

\noindent
From (\ref{massWL}) and (\ref{massWH}) we find
\be\label{ratioMHML}
M_{W_H^\pm}=\frac{f}{v}\frac{M_{W_L^\pm}}{sc}, 
\ee
which is valid to order $v^2/f^2$.

The formulae given above have been already presented in \cite{Logan}
but at a few places our results differ from the ones presented there. 
We would like to spell out these differences explicitly.
\begin{itemize}
\item
 In going from 
(\ref{Lmasses}) to (\ref{eigenvectors}) we have not made any field 
redefinitions as
done in \cite{Logan}.  As a result of this, the formulae in  
(\ref{eigenvectors}) differ from (A34) in \cite{Logan} by $B$ replaced by
$-B$. This difference is a matter of choice and has no impact 
on physical results.
\item
Our results for $x_Z^{W'}$ and $x_Z^{B'}$ in (\ref{xWxB})
differ by signs from the ones given in (A35) of \cite{Logan}. 
This difference is crucial for the removal of the divergences in our
calculations in the unitary gauge.
\item
As seen in (\ref{MAH}) and (\ref{MZH}) we do not confirm at this order
the presence of terms proportional to $x_H$ that can be found in (A37) of
\cite{Logan}. Our result is consistent with the LH model with T-parity
\cite{Hubiszpheno}-\cite{BBPTUW2}, where the terms proportional to $x_H$ are also absent at this order.
\end{itemize}

\subsection{The Fermion Sector}
Here it suffices to state that in addition to the standard quarks and leptons
there is a new heavy top quark $T$ with the mass

\be \label{mT}
m_T=\frac{f}{v}\frac{m_t}{\sqrt{x_L (1-x_L)}}\left(1+\frac{v^2}{f^2} \left(\frac{1}{3}-x_L(1-x_L)\right)\right),
\qquad
x_L=\frac{\lambda_1^2}{\lambda_1^2+\lambda_2^2},
\ee
where $\lambda_1$ is the Yukawa coupling in the $(t,T)$ sector and
$\lambda_2$  parametrizes the mass term for $T$. 
As already discussed in \cite{Logan,BPU04,DEC05}, the
parameter $x_L$ describes together with $v/f$ the size of the violation of
the three generation CKM unitarity and is also crucial for the gauge
interactions of the heavy $T$ quark with the ordinary down quarks. 
$\lambda_i$ are expected to be $\ord(1)$ with \cite{Logan} 
\be
\lambda_i\ge \frac{m_t}{v}, \qquad {\rm or}\qquad
\frac{1}{\lambda_1^2}+ \frac{1}{\lambda_2^2}\approx
\left(\frac{v}{m_t}\right)^2
\ee
so that within a good approximation
\be
\lambda_1=\frac{m_t}{v}\frac{1}{\sqrt{1-x_L}}, \qquad 
\lambda_2=\frac{m_t}{v}\frac{1}{\sqrt{x_L}}.
\ee
$x_L$ can in principle vary in the range $0<x_L<1$. For $x_L\approx 0$ and
$x_L\approx 1$, the mass $m_T$ becomes very large \cite{DEC05}.
\subsection{The Charged Scalar Sector}
The LH model contains also a triplet of heavy scalars of which only
$\Phi^\pm$ will be of relevance here. In the case of particle-antiparticle
mixing and box diagram contributions to rare decays, $\Phi^\pm$ do not
contribute at $\ord(v^2/f^2)$, but they contribute at this order to 
$Z_L$-penguin diagrams. The mass of $\Phi^\pm$ is given by
\be
M_{\Phi^\pm}=\sqrt{2} m_H\frac{f}{v}
\ee
with $m_H$ denoting the light Higgs mass. 

\subsection{Feynman Rules}
\subsubsection{Charged Gauge Boson--Fermion Interactions}
The Feynman rules for vertices involving the charged $W_{L}^\pm$ and
$W_{H}^\pm$ bosons and quarks in the notation $C
  \gamma_{\mu} (1-\gamma_5)$ are given in Table~\ref{tab:feyn} where
\be\label{defa}
a=\frac{1}{2}c^2(c^2-s^2), \qquad b=\frac{1}{2}s^2(c^2-s^2),\qquad
d_2=-\frac{5}{6}+\frac{1}{2} x_L^2+2 x_L (1-x_L).
\ee
 $x_L$ is given in (\ref{mT}).
For leptons the Feynman rules
can be obtained from the entries of the first line with $V_{ij}=1$. The
$V_{ij}$ are the usual CKM parameters. The issue of the violation 
of the CKM unitarity at $\ord(v^2/f^2)$ has been already discussed in detail
in \cite{BPU04} and will not be repeated here.  Table~\ref{tab:feyn} 
should be compared with Table VIII of \cite{Logan}.
 Due to different phase conventions for the $t$
and $T$ fields, our rules for the vertices $W_L^\pm\bar T d_j$ and 
$W_H^\pm\bar T d_j$ differ by a crucial factor $i$ as already 
discussed in \cite{BPU04}.
\begin{table}[htb]
\caption[b]{Feynman Rules in Littlest Higgs Model for $W_{L,H}$.
 $C
  \gamma_{\mu} (1-\gamma_5)$.
\label{tab:feyn}}
\begin{center}
\begin{tabular}{|l|l|l|l|}\hline
Vertex&$ C$& Vertex&$C$\\
\hline
$W_L^{+} \bar u_i d_j$& $\frac{i g}{2\sqrt{2}}\;V_{ij}\left(1-\frac{v^2}{f^2}a\right)$&$W_H^{+}\bar u_id_j$&$-\frac{i
  g}{2\sqrt{2}}\;V_{ij}\frac{c}{s}
\;\left(1+b\frac{v^2}{f^2}\right)$\\
\hline
$W_L^{+}\bar td_j$&$\frac{i g}{2\sqrt{2}}\;V_{tj}\;
\left(1-\frac{v^2}{f^2}(\frac{1}{2} x^2_L+a)\right)$&$W_H^{+}\bar td_j$&$-\frac{i
  g}{2\sqrt{2}}\;V_{tj}\frac{c}{s}
\;\Big(1-\frac{v^2}{f^2}(\frac{1}{2} x^2_L-b)\Big)$\\
\hline
$W_L^{+}\bar Td_j$&$\frac{i
  g}{2\sqrt{2}}\;V_{tj}\;x_L\;\frac{v}{f}\left(1+\frac{v^2}{f^2}\left(d_2-a\right)\right)$&$W_H^{+}\bar Td_j$&$-\frac{i
  g}{2\sqrt{2}}\;V_{tj}\frac{c}{s}
\;x_L\;\frac{v}{f}$\\
\hline
\end{tabular}
\end{center}
\end{table}

\subsubsection{Neutral Gauge Boson--Fermion Interactions}
The vertices involving quarks and leptons and the neutral gauge
bosons $Z^{0}_{L}$, $Z^{0}_{H}$ and $A^{0}_{H}$, that are
relevant for our paper, are presented in
Table~\ref{tab:feyn2}, where $g_V$ and $g_A$
parametrize universally the vertices as follows
\be\label{gVgA}
i\gamma_\mu(g_V+g_A\gamma_5)
\ee
and
\be\label{defu}
u=(c'^2-s'^2),\qquad a'=\frac{1}{2}c'^2 (c'^2-s'^2).
\ee
These rules follow from (A55) of \cite{Logan} that we confirmed except 
for the signs in $x_Z^{W'}$ and $x_Z^{B'}$ in (\ref{xWxB}) as discussed 
above. In spite of agreeing with (A55) the rules presented in 
Table~\ref{tab:feyn2} differ surprisingly at various places from Table 
IX of \cite{Logan}. The differences are found in the couplings 
$Z_L\bar u u$, $Z_L\bar t t$, $Z_L\bar T t$, $A_H\bar T T$ and $Z_H\bar TT$.
 They all are crucial for 
the cancellation of the divergences in our calculation. 
In order to make the comparison 
with \cite{Logan} as simple as possible, 
Table~\ref{tab:feyn2} has exactly the same form as the table IX of 
\cite{Logan}. Table~\ref{tab:feyn2} contains also higher order terms 
in $v/f$ that were required in our calculation of diagrams in classes 
4 and 5 discussed below and were not present in \cite{Logan}. 

As discussed in \cite{Logan}, the gauge invariance of the
Yukawa interactions alone cannot unambiguously fix all the $U(1)$ charge
values. The two parameters $y_e$ and $y_u$ that enter the Feynman Rules in
Table~\ref{tab:feyn2} are undetermined. If one requires that the $U(1)$ charge
assignments be anomaly free, they can be fixed to be
\be\label{anomf}
y_e=\frac{3}{5}, \qquad  y_u=-\frac{2}{5}.
\ee
On the other hand, as emphasized in \cite{Logan}, in an effective field
theory below a cutoff, it is unnecessary to be completely anomaly free as the
anomalies could be cancelled by some specific extra matter at the cutoff
scale. In the rest of the paper we will set $y_e$ and $y_u$ to the values
given in (\ref{anomf}) in order to avoid additional sensitivity to the
physics at the cut-off scale.

\begin{table}
\caption[]{Feynman Rules in LH Model for $Z_{L}$, $A_H$
  and $Z_H$. $g_V$ and $g_A$ defined in (\ref{gVgA}). \label{tab:feyn2}} 
\begin{tabular}{|c|c|c|}
\hline
\!\!\!\!\!\!\!\!{\small vertex}\!\!\!\!\!\!\!\! & $g_V$ & $g_A$ \\
\hline
$\!\!\!\!A_L \bar f f\!\!$ & $-e Q_f$ & $0$ \\
\hline
$\!\!\!\!Z_L \bar u u\!\!$ & 
        $-\frac{g}{2c_{w}} \left\{ (\frac{1}{2} - \frac{4}{3} s^2_{w})
        - \frac{v^2}{f^2} \left[ c_{w} x_Z^{W^{\prime}} c/2s
        \right. \right.$ &
        $-\frac{g}{2c_{w}} \left\{ -\frac{1}{2}
        - \frac{v^2}{f^2} \left[ -c_{w} x_Z^{W^{\prime}} c/2s
        \right. \right.$ \\
$$ &
        $\left. \left.
        + \frac{s_{w} x_Z^{B^{\prime}}}{s^{\prime}c^{\prime}}
        \left( 2y_u + \frac{17}{15} - \frac{5}{6} c^{\prime 2}
        \right) \right] \right\} $
        &
        $\left. \left.
        + \frac{s_{w} x_Z^{B^{\prime}}}{s^{\prime}c^{\prime}}
        \left( \frac{1}{5} - \frac{1}{2} c^{\prime 2} \right) 
        \right] \right\}$ \\
\hline
$\!\!\!\!Z_L \bar d d\!\!$ &
        $-\frac{g}{2c_{w}} \left\{ (-\frac{1}{2} + \frac{2}{3} s^2_{w})
        - \frac{v^2}{f^2} \left[ -c_{w} x_Z^{W^{\prime}} c/2s
        \right. \right.$ &
        $-\frac{g}{2c_{w}} \left\{ \frac{1}{2}
        - \frac{v^2}{f^2} \left[ c_{w} x_Z^{W^{\prime}} c/2s
        \right. \right.$ \\
$$ &
        $\left. \left.
        + \frac{s_{w} x_Z^{B^{\prime}}}{s^{\prime}c^{\prime}}
        \left( 2y_u + \frac{11}{15} + \frac{1}{6} c^{\prime 2}
        \right) \right] \right\}$ &
        $\left. \left.
        + \frac{s_{w} x_Z^{B^{\prime}}}{s^{\prime}c^{\prime}}
        \left( -\frac{1}{5} + \frac{1}{2} c^{\prime 2} \right)
        \right] \right\}$ \\
\hline
$\!\!\!\!Z_L \bar e e\!\!$ &
        $-\frac{g}{2c_{w}} \left\{ (-\frac{1}{2} + 2 s^2_{w})
        - \frac{v^2}{f^2} \left[ -c_{w} x_Z^{W^{\prime}} c/2s
        \right. \right.$ &
        $-\frac{g}{2c_{w}} \left\{ \frac{1}{2}
        - \frac{v^2}{f^2} \left[ c_{w} x_Z^{W^{\prime}} c/2s
        \right. \right.$ \\
$$ &
        $\left. \left.
        + \frac{s_{w} x_Z^{B^{\prime}}}{s^{\prime}c^{\prime}}
        \left( 2y_e - \frac{9}{5} + \frac{3}{2} c^{\prime 2} 
        \right) \right] \right\}$ &
        $\left. \left.
        + \frac{s_{w} x_Z^{B^{\prime}}}{s^{\prime}c^{\prime}}
        \left( -\frac{1}{5} + \frac{1}{2} c^{\prime 2} \right)
        \right] \right\}$ \\
\hline
$\!\!\!\!Z_L \bar \nu \nu\!\!$ &
        $-\frac{g}{2c_{w}} \left\{ \frac{1}{2}
        - \frac{v^2}{f^2} \left[ c_{w} x_Z^{W^{\prime}} c/2s
        \right. \right.$ &
        $-\frac{g}{2c_{w}} \left\{ -\frac{1}{2}
        - \frac{v^2}{f^2} \left[ -c_{w} x_Z^{W^{\prime}} c/2s
        \right. \right.$ \\
$$ &
        $
        + \frac{s_{w} x_Z^{B^{\prime}}}{s^{\prime}c^{\prime}}
        \left( y_e - \frac{4}{5} + \frac{1}{2} c^{\prime 2}
        \right) \Big] \Big\}$ &
        $\left. \left.
        + \frac{s_{w} x_Z^{B^{\prime}}}{s^{\prime}c^{\prime}}
        \left( -y_e + \frac{4}{5} - \frac{1}{2} c^{\prime 2}
        \right) \right] \right\}$ \\
\hline
$\!\!\!\!\!\!Z_L \bar t t\!\!$ &
        $-\frac{g}{2c_{w}} \Big\{ (\frac{1}{2} - \frac{4}{3} s^2_{w})
        - \frac{v^2}{f^2} \left[  x_L^2/2 
        + c_{w} x_Z^{W^{\prime}} c/2s
        \right. $ &
        $-\frac{g}{2c_{w}} \left\{ -\frac{1}{2}
        - \frac{v^2}{f^2} \left[ -x_L^2/2
        - c_{w} x_Z^{W^{\prime}} c/2s
        \right. \right.$ \\
$$ &
        $
        + \frac{s_{w} x_Z^{B^{\prime}}}{s^{\prime}c^{\prime}}
        \left( 2y_u + \frac{17}{15} - \frac{5}{6} c^{\prime 2}
        - \frac{1}{5} 
        \frac{\lambda_1^2}{\lambda_1^2 + \lambda_2^2}
        \right) \Big] \Big\}$ &
        $\left. \left.
        + \frac{s_{w} x_Z^{B^{\prime}}}{s^{\prime}c^{\prime}}
        \Big( \frac{1}{5} - \frac{1}{2} c^{\prime 2}
        - \frac{1}{5} \frac{\lambda_1^2}{\lambda_1^2 + \lambda_2^2}
        \Big) \right] \right\}$ \\
\hline
$\!\!Z_L \bar T T\!\!$ &
        $\frac{g}{2
          c_w}\Big\{\frac{4}{3}s_w^2+\frac{v^2}{f^2}\Big(-
\frac{1}{2}x_L^2+$ &
        $\frac{g}{2
          c_w}\frac{v^2}{f^2}\Big\{
\frac{1}{2}x_L^2+\frac{s_w x_Z^{B'}}{s' c'}\frac{1}{5}x_L\Big\}$ \\
&$\frac{s_w x_Z^{B'}}{s' c'}(2 y_u+\frac{14}{15}-\frac{4}{3}c'^2+\frac{1}{5}x_L)\Big)\Big\}$&\\
\hline
$\!\!\!\!Z_L \bar T t\!\!$ &
        $\!\!\!\!\!\!\!\frac{g}{2 c_w}\Big\{\!\!-\!\frac{v}{f}\frac{1}{2}
            x_L\!+\!\frac{v^2}{f^2}\frac{s_w x_Z^{B'}}{c'
              s'}\left(\frac{1}{5} x_L \frac{\lambda_2}{\lambda_1}\right)\!\!+\!\!\!\!$ &
        $\!\!\!\!\frac{g}{2 c_w}\Big\{\frac{v}{f}\frac{1}{2}
            x_L\!+\!\frac{v^2}{f^2}\frac{s_w x_Z^{B'}}{c'
              s'}\left(\frac{1}{5} x_L \frac{\lambda_2}{\lambda_1}\right)\!+\!\!\!\!$ \\
&$\!\!\!\!\!\!\!\frac{v^3}{f^3}\Big(\!\frac{1}{4}x_L^3\!-\!\frac{1}{2}x_L d_2\!+\!x_L
  (\frac{c'}{s'}\frac{s_w x_Z^{B'}}{2}\!+\!\frac{c}{s}\frac{c_w x_Z^{W'}}{2})\Big)\!\Big\}\!\!\!\!\!$&$\!\!\frac{v^3}{f^3}\Big(\!\!\!-\!\frac{1}{4}x_L^3\!+\!\frac{1}{2}x_L d_2\!-\!x_L
  (\!\frac{c'}{s'}\frac{s_w x_Z^{B'}}{2}\!+\!\frac{c}{s}\frac{c_w x_Z^{W'}}{2})\Big)\!\Big\}\!\!\!$\\
\hline
$\!\!\!\!A_H \bar u u\!\!$ & 
        $\frac{g^{\prime}}{2s^{\prime}c^{\prime}}
        \left( 2y_u + \frac{17}{15} - \frac{5}{6} c^{\prime 2} \right)$ &
        $\frac{g^{\prime}}{2s^{\prime}c^{\prime}}
        \left( \frac{1}{5} - \frac{1}{2} c^{\prime 2} \right)$ \\
$\!\!\!\!A_H \bar d d\!\!$ &
        $\frac{g^{\prime}}{2s^{\prime}c^{\prime}}
        \left( 2y_u + \frac{11}{15} + \frac{1}{6} c^{\prime 2} \right)$ &
        $\frac{g^{\prime}}{2s^{\prime}c^{\prime}}
        \left( -\frac{1}{5} + \frac{1}{2} c^{\prime 2} \right)$ \\
$\!\!\!\!A_H \bar e e\!\!$ &
        $\frac{g^{\prime}}{2s^{\prime}c^{\prime}}
        \left( 2y_e - \frac{9}{5} + \frac{3}{2} c^{\prime 2} \right)$ &
        $\frac{g^{\prime}}{2s^{\prime}c^{\prime}}
        \left( -\frac{1}{5} + \frac{1}{2} c^{\prime 2} \right)$ \\
$\!\!\!\!A_H \bar \nu \nu\!\!$ &
        $\frac{g^{\prime}}{2s^{\prime}c^{\prime}}
        \left( y_e - \frac{4}{5} + \frac{1}{2} c^{\prime 2} \right)$ &
        $\frac{g^{\prime}}{2s^{\prime}c^{\prime}}
        \left( -y_e + \frac{4}{5} - \frac{1}{2} c^{\prime 2} \right)$ \\
\hline
$\!\!\!\!A_H \bar t t\!\!$ &
        $\frac{g^{\prime}}{2s^{\prime}c^{\prime}}
        \left( 2y_u + \frac{17}{15} - \frac{5}{6} c^{\prime 2}
        - \frac{1}{5} \frac{\lambda_1^2}{\lambda_1^2 + \lambda_2^2} \right)$ &
        $\frac{g^{\prime}}{2s^{\prime}c^{\prime}}
        \left( \frac{1}{5} - \frac{1}{2} c^{\prime 2}
        - \frac{1}{5} \frac{\lambda_1^2}{\lambda_1^2 + \lambda_2^2} \right)$ \\
$\!\!A_H \bar T T\!\!$ &
        $\frac{g^{\prime}}{2s^{\prime}c^{\prime}}
        \left( 2y_u + \frac{14}{15} - \frac{4}{3} c^{\prime 2}
        + \frac{1}{5} \frac{\lambda_1^2}{\lambda_1^2 + \lambda_2^2} \right)$ &
        $\frac{g^{\prime}}{2s^{\prime}c^{\prime}}
        \frac{1}{5} \frac{\lambda_1^2}{\lambda_1^2 + \lambda_2^2}$ \\
$\!\!\!\!A_H \bar T t\!\!$ &
        $\frac{g^{\prime}}{2s^{\prime}c^{\prime}}
        \Big(\frac{1}{5}
        x_L\frac{\lambda_2}{\lambda_1}+\frac{v}{f}\frac{1}{2}c'^2 x_L\Big)$ &
        $\frac{g^{\prime}}{2s^{\prime}c^{\prime}}
        \Big(\frac{1}{5}
        x_L\frac{\lambda_2}{\lambda_1}-\frac{v}{f}\frac{1}{2}c'^2 x_L\Big)$ \\
\hline
$\!\!\!\!Z_H \bar u u\!\!$ & $gc/4s$ & $-gc/4s$ \\
$\!\!\!\!Z_H \bar d d\!\!$ & $-gc/4s$ & $gc/4s$ \\
$\!\!\!\!Z_H \bar e e\!\!$ & $-gc/4s$ & $gc/4s$ \\
$\!\!\!\!Z_H \bar \nu \nu\!\!$ & $gc/4s$ & $-gc/4s$ \\
\hline
$\!\!Z_H \bar t t\!\!$ & $gc/4s$ & $-gc/4s$ \\
$\!\!Z_H \bar T T\!\!$ & $\mathcal{O}(v^2/f^2)$ & $\mathcal{O}(v^2/f^2)$ \\
$\!\!\!\!Z_H \bar T t\!\!$ &
        $ g x_L vc/4fs$ &
        $- g x_L vc/4fs$ \\
\hline
\end{tabular}
\end{table}

We do not present the rules for the triple gauge boson vertices as they can
be found in Table VII of \cite{Logan}. 

\subsubsection{Charged Scalar Interactions}
Only the following Feynman rules given in \cite{Logan} are of relevance in the
present paper:
\be
\Phi^+\bar u_i d_j:\quad -\frac{i}{\sqrt{2}}\frac{g}{4}
                           \frac{m_i}{M_{W_L}}(1-\gamma_5)\frac{v}{f}V_{ij}
\ee
\be
\Phi^+\bar T d_j:\quad -\frac{i}{\sqrt{2}}\frac{g}{4}
                           \frac{m_t}{M_{W_L}}(1-\gamma_5)
                    \frac{\lambda_1}{\lambda_2}  \frac{v}{f}V_{tj}
\ee
\be
\Phi^+\Phi^- Z_L:\quad i\frac{g}{c_w}s^2_w(p_+-p_-)_\mu
\ee
with $p_\pm$ being outgoing momenta of $\Phi^\pm$. 
For the $\Phi^-\bar d_j u_i$ vertex $(1-\gamma_5)$ should replaced by 
$(1+\gamma_5)$ and $V_{ij}$ by $V_{ij}^*$. Similarly for $\Phi^-\bar d_j
T$. 
\boldmath
\section{$X$ and $Y$ in the Standard Model}\label{sec:XYSM}
\unboldmath
\setcounter{equation}{0}
Many rare decays in the SM are governed by the functions $X(x_t)$ and
$Y(x_t)$ with $x_t=m_t^2/M_W^2$. 
It will turn out to be useful to recall the structure of the
calculation of these functions. Calculating the $Z^0$-penguin contribution to
the effective Hamiltonian for decays with $\nu\bar\nu$ and $\mu\bar\mu$ in
the final state one finds
\be\label{ZHnunu}
({\cal{H}}^{\nu\bar\nu}_{\rm eff})_Z=\frac{g^4}{64\pi^2}
\frac{1}{M_Z^2\cos^2\theta_w} C(x_t)(\bar s d)_{V-A}(\bar\nu\nu)_{V-A},
\ee
\be\label{ZHmumu}
({\cal{H}}^{\mu\bar\mu}_{\rm eff})_Z=-\frac{g^4}{64\pi^2}
\frac{1}{M_Z^2\cos^2\theta_w} C(x_t)(\bar s d)_{V-A}(\bar\mu\mu)_{V-A}. 
\ee
The corresponding calculation of the box diagrams gives
\be\label{BHnunu}
({\cal{H}}^{\nu\bar\nu}_{\rm eff})_{\rm Box}=\frac{g^4}{64\pi^2}
\frac{1}{M_W^2} B^{\nu\bar\nu}(x_t) (\bar s d)_{V-A}(\bar\nu\nu)_{V-A},
\ee
\be\label{BHmumu}
({\cal{H}}^{\mu\bar\mu}_{\rm eff})_{\rm Box}=-\frac{g^4}{64\pi^2}
\frac{1}{M_W^2} B^{\mu\bar\mu}(x_t) (\bar s d)_{V-A}(\bar\mu\mu)_{V-A}. 
\ee
Adding the $Z^0$-penguin and box contributions and
using the relations
\be\label{rel}
M_Z^2\cos^2\theta_w=M_W^2,\qquad \frac{G_F}{\sqrt{2}}=\frac{g^2}{8 M^2_W}
\ee
one arrives at
\be\label{Hnunu}
{\cal{H}}^{\nu\bar\nu}_{\rm eff}= M_W^2\frac{G_F^2}{2\pi^2}
 X(x_t)(\bar s d)_{V-A}(\bar\nu\nu)_{V-A},
\ee
\be\label{Hmumu}
{\cal{H}}^{\mu\bar\mu}_{\rm eff}=-M_W^2\frac{G_F^2}{2\pi^2}
Y(x_t)(\bar s d)_{V-A}(\bar\mu\mu)_{V-A}, 
\ee
where
\be\label{XY}
X(x_t)=C(x_t)+B^{\nu\bar\nu}(x_t), \qquad
Y(x_t)=C(x_t)+B^{\mu\bar\mu}(x_t).
\ee
It is customary to use in (\ref{Hnunu}) and (\ref{Hmumu}) the relation
\be
M_W^2\frac{G_F^2}{2\pi^2}=
\frac{G_F}{\sqrt{2}}\frac{\alpha}{2\pi\sin^2\theta_w},
\ee
but we will not use it here for reasons discussed in the next section.

Now, $C(x_t)$, $B^{\nu\bar\nu}(x_t)$ and $B^{\mu\bar\mu}(x_t)$ depend on 
the gauge used for the $W^\pm$ propagator. One has \cite{IL,PBE}
\be
C(x_t)=C_0(x_t)+\frac{1}{2}\bar\varrho(x_t)
\ee
\be
B^{\nu\bar\nu}(x_t)=-4 B_0(x_t)-\frac{1}{2}\bar\varrho(x_t), 
\qquad
B^{\mu\bar\mu}(x_t)=- B_0(x_t)-\frac{1}{2}\bar\varrho(x_t),
\ee 
where $\bar\varrho(x_t)$ is gauge dependent with $\bar\varrho(x_t)=0$ in the 
Feynman--t'Hooft gauge and
\begin{equation}\label{BF}
B_0(x_t)=\frac{1}{4}\left[\frac{x_t}{1-x_t}+\frac{x_t\log x_t}{(x_t-1)^2}\right],
\end{equation}
\begin{equation}\label{C0xt}
C_0(x_t)=\frac{x_t}{8}\left[\frac{x_t-6}{x_t-1}+\frac{3x_t+2}{(x_t-1)^2}\;\log x_t\right]~.
\end{equation}

Evidently $X(x_t)$ and $Y(x_t)$ are gauge independent and given  
in the SM as follows:
\begin{equation}\label{X0}
X(x_t)=\frac{x_t}{8}\;\left[\frac{x_t+2}{x_t-1} 
+\frac{3 x_t-6}{(x_t -1)^2}\; \log x_t\right], 
\end{equation}
\begin{equation}\label{Y0}
Y(x_t)=\frac{x_t}{8}\; \left[\frac{x_t -4}{x_t-1} 
+ \frac{3 x_t}{(x_t -1)^2} \log x_t\right].
\end{equation}

Explicit expression for $\bar\varrho(x_t)$ in an arbitrary $R_\xi$ gauge 
can be found in \cite{IL,PBE}. In the LH model we will calculate $X$ and $Y$ 
in the unitary gauge and we will need at one stage   
the SM function $C(x_t)$ in this gauge. As the penguin and box diagrams are
divergent in the unitary gauge, even after the GIM mechanism has been 
invoked, we 
use the dimensional regularization with $D=4-2\varepsilon$ to 
find
\be\label{Cunit}
C(x_t)_{\rm unitary}=-\frac{1}{16} x_t \left(\frac{1}{\varepsilon} +
\ln \frac{\mu^{2}}{M_{W_{L}}^{2}}\right) -
\frac{x_t^2-7 x_t}{32(1-x_t)} +
\frac{4 x_t-2 x_t^2+x_t^3}{16(1-x_t)^2} \log x_t
\ee
and 
\be\label{Runit}
\bar{\varrho}(x_t)_{\rm unitary}=-\frac{1}{8} x_t \left(\frac{1}{\varepsilon} +
\ln \frac{\mu^{2}}{M_{W_{L}}^{2}}\right) -
\frac{-3 x_t^2 + 17 x_t}{16(1-x_t)}-
\frac{8 x_t^2 - x_t^3}{8(x_t-1)^2} \log x_t~.
\ee
The $\ln \left(\mu^{2}/M_{W_{L}}^{2}\right)$ terms disappear in the final
expressions for $X$ and $Y$ as they always accompany $1/\varepsilon$ that
is not present in $X$ and $Y$ in the SM.
\boldmath
\section{$X$ and $Y$ in the Littlest Higgs Model}\label{sec:XYLH}
\unboldmath
\setcounter{equation}{0}
\subsection{Six Classes of Diagrams}

\begin{figure}[ht]
\vspace{0.10in}
\epsfysize=3.5in
\centerline{
\epsffile{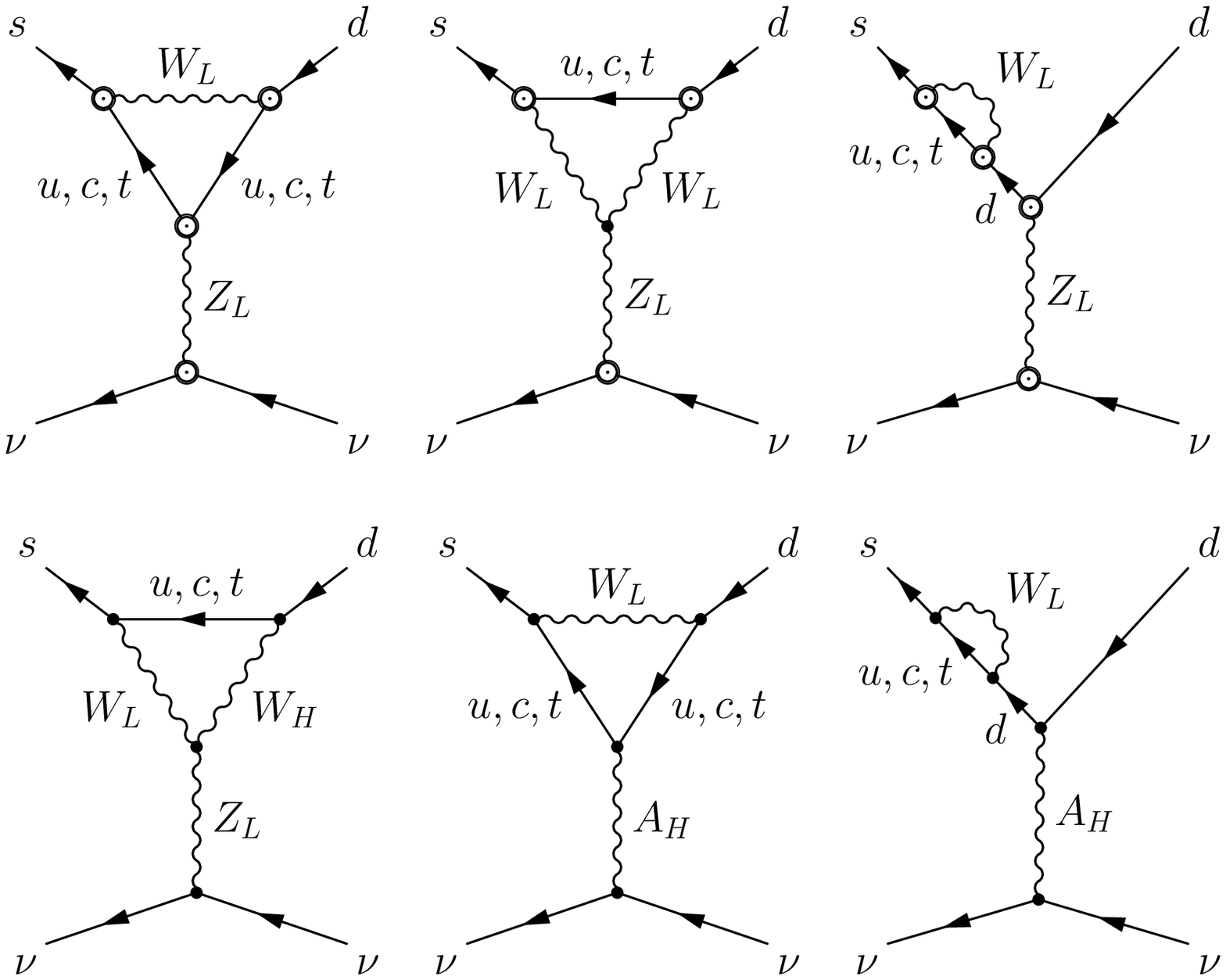}}
\vspace{0.01in}
\caption{Class 1. Penguin and box diagrams with SM particles and $A_{H}$
contributing to $K\to \pi\nu\bar\nu$ within the LH model
at $\ord(v^2/f^2)$.}\label{fig:class1}
\end{figure}
\begin{figure}[ht]
\vspace{0.10in}
\epsfysize=3.5in
\centerline{
\epsffile{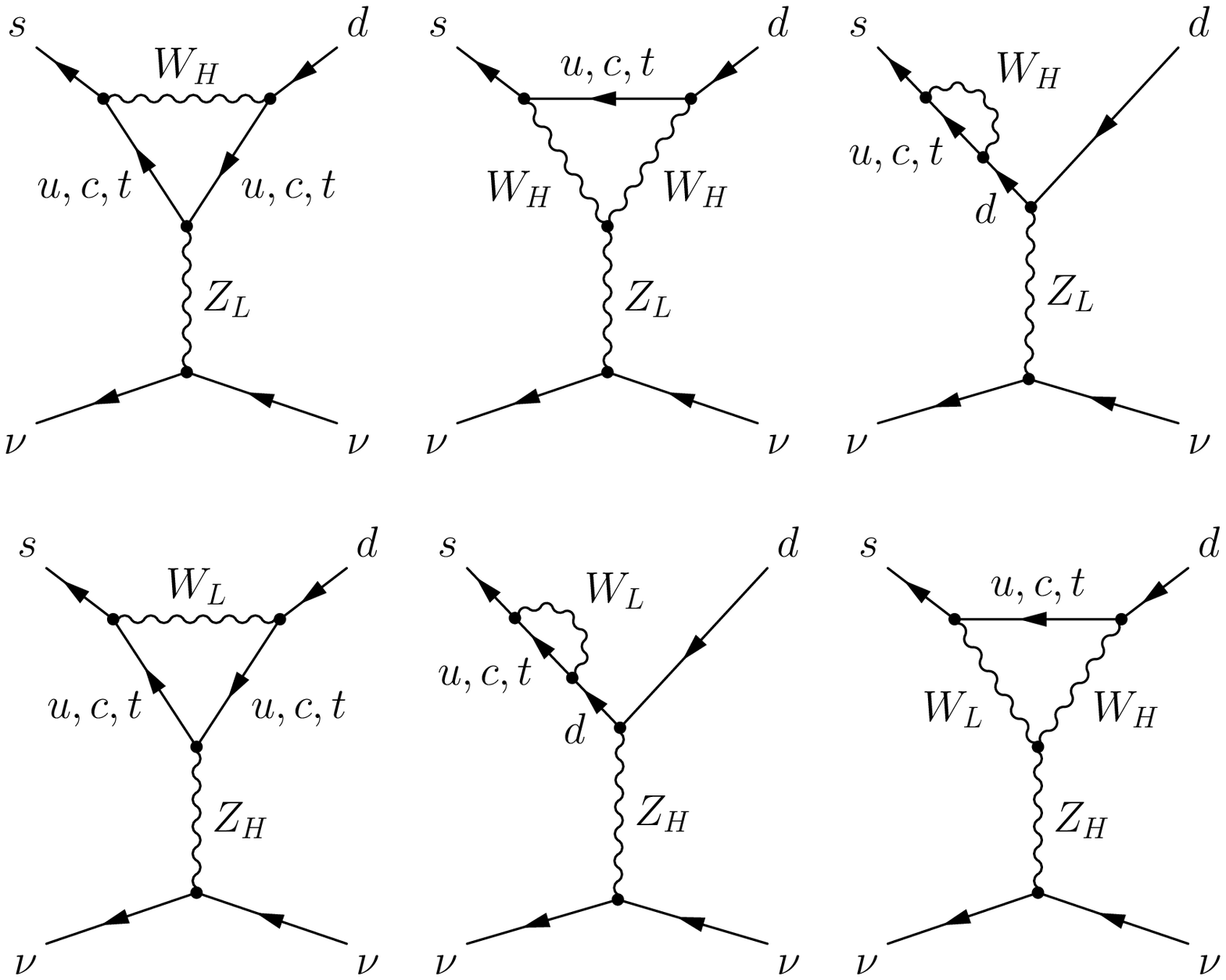}}
\vspace{0.01in}
\caption{Class 2. Penguin and box diagrams with $W_{H}$ and $Z_{H}$
contributing to $K\to \pi \nu\bar\nu$ within the LH model
at $\ord(v^2/f^2)$.}\label{fig:class2}
\end{figure}
\begin{figure}[ht]
\vspace{0.10in}
\epsfysize=5.25in
\centerline{
\epsffile{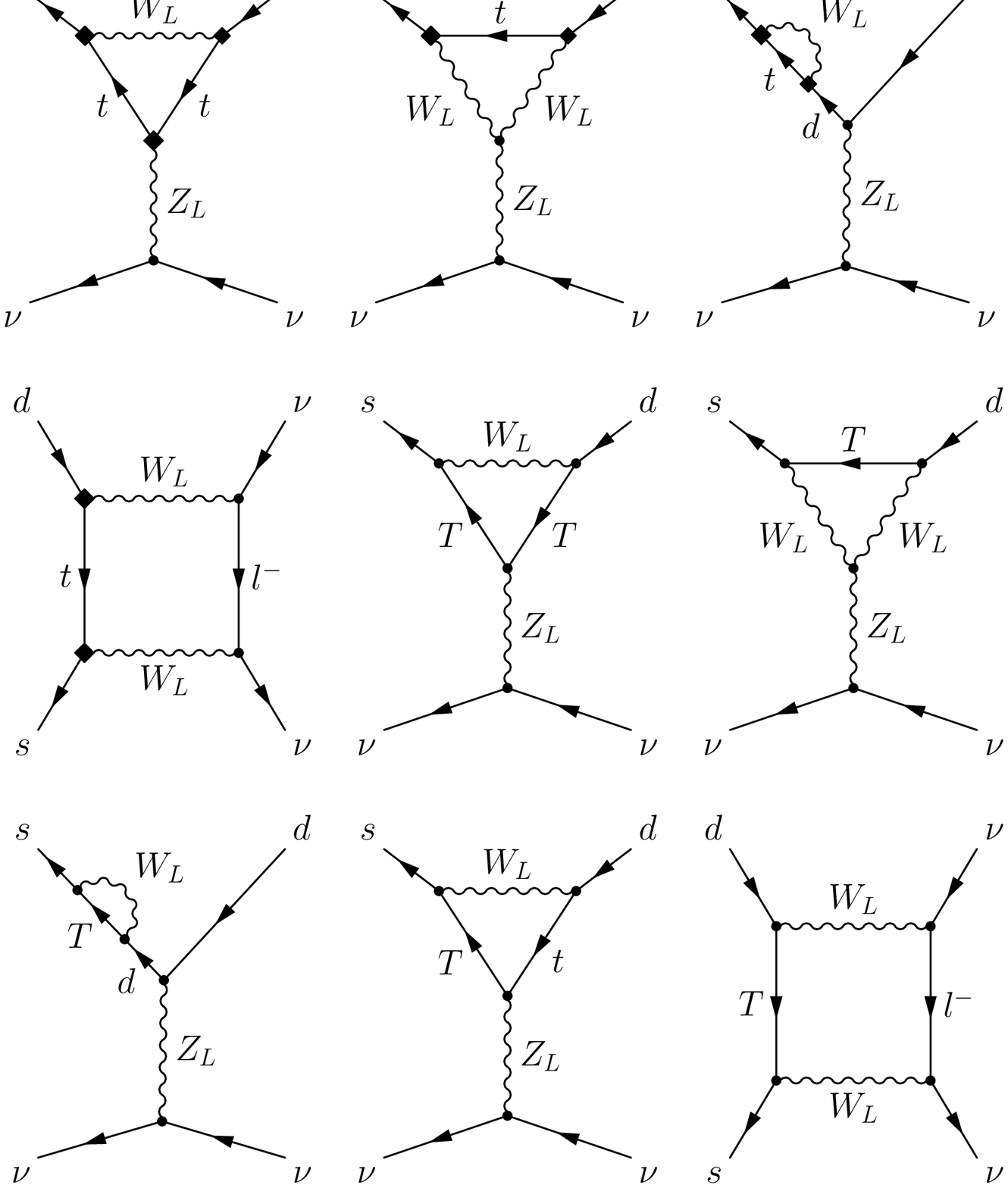}}
\vspace{0.01in}
\caption{Class 3. Top and heavy top quark contributions 
to $K \to\pi \nu\bar\nu$ in
  the LH model at $\ord (v^2/f^2)$ which are proportional to $x_L^2$.}\label{fig:class3}
\end{figure}

\begin{figure}[ht]
\vspace{0.10in}
\epsfysize=3.5in
\centerline{
\epsffile{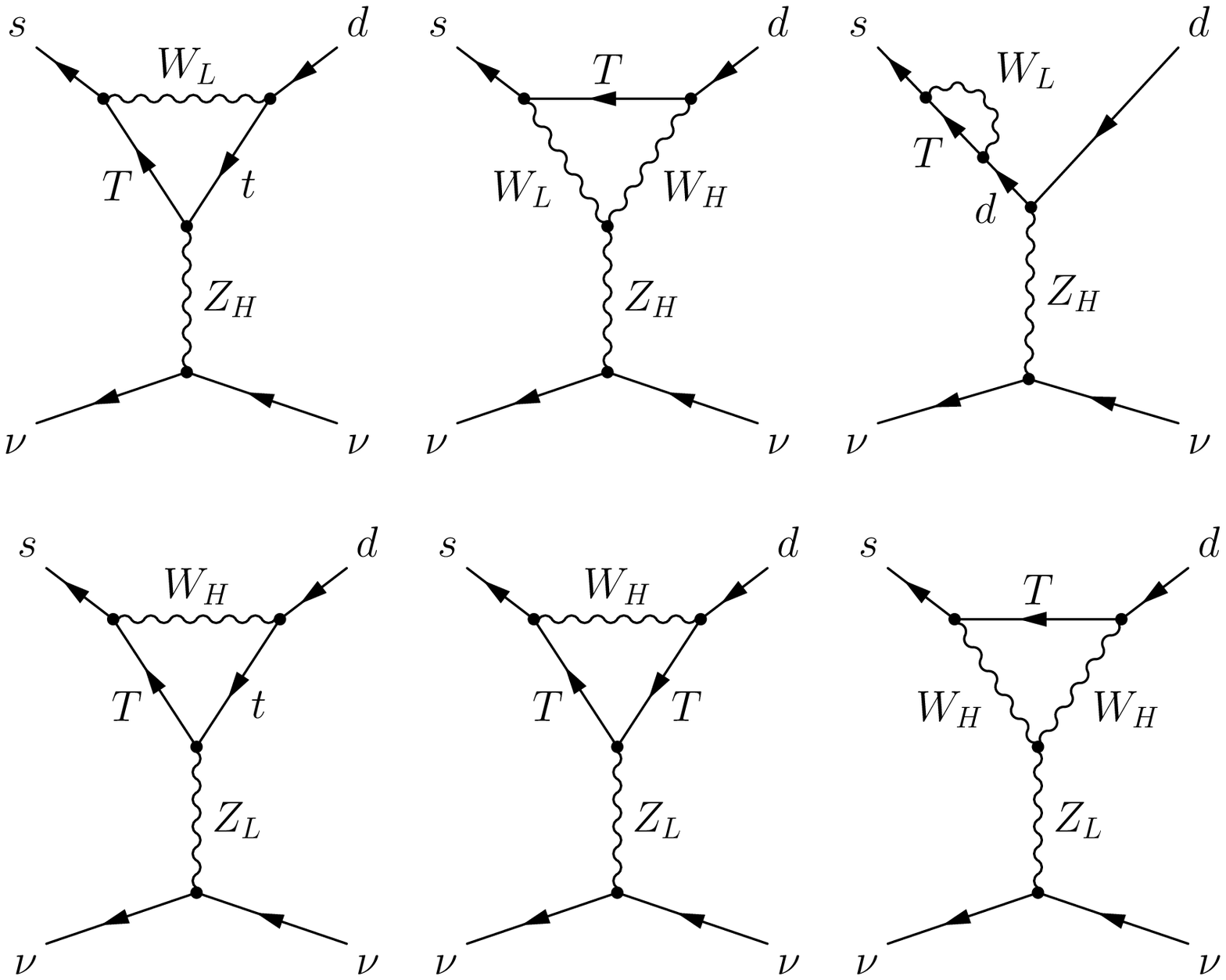}}
\vspace{0.01in}
\caption{Class 4. Penguin and box contributions
to $K \to\pi \nu\bar\nu$ in
  the LH model at $\ord (v^2/f^2)$ which are proportional to $v^4/f^4 c^4
x_L^2$.}\label{fig:class4}
\end{figure}

\begin{figure}[ht]
\vspace{0.10in}
\epsfysize=5.25in
\centerline{
\epsffile{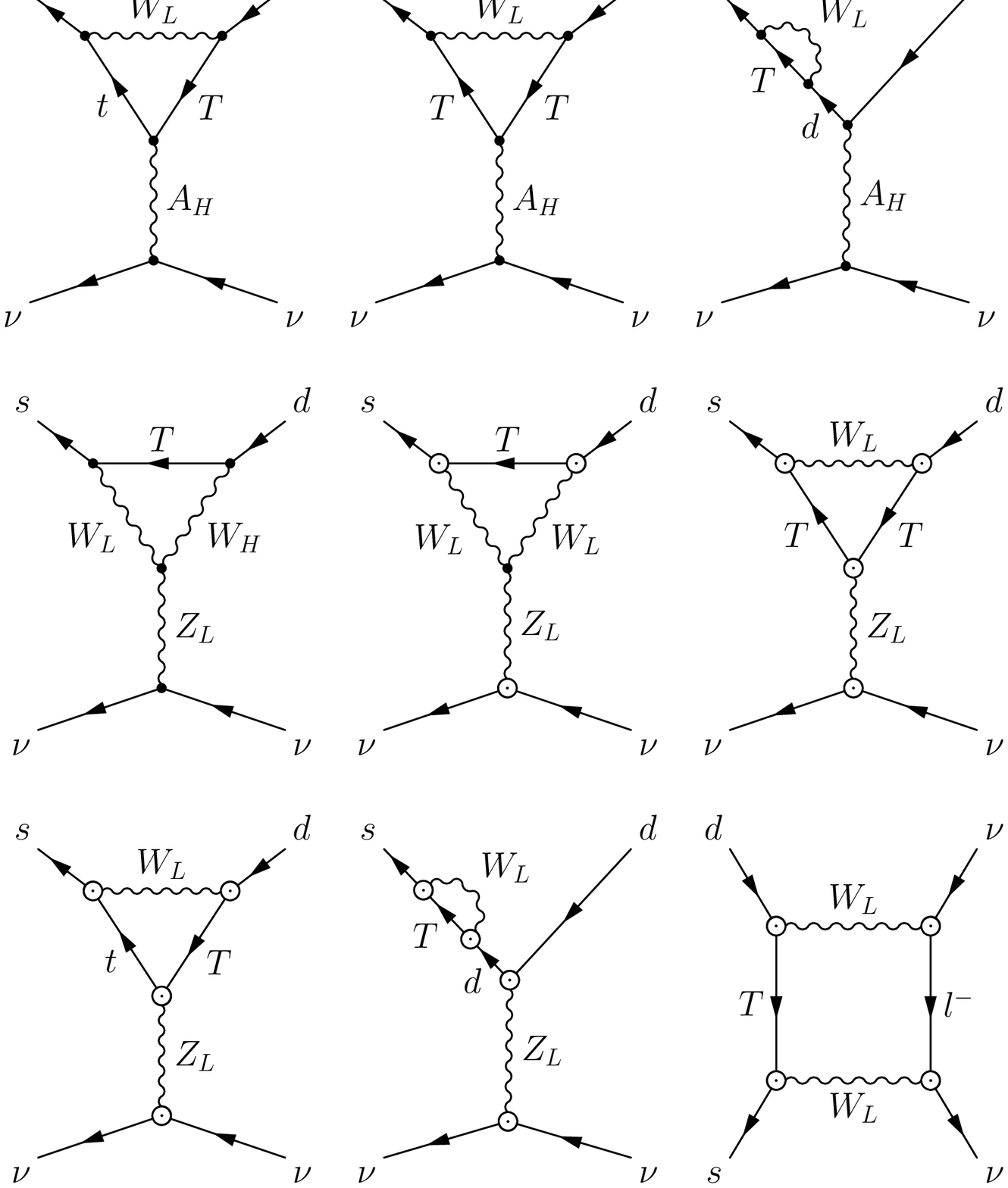}}
\vspace{0.01in}
\caption{Class 5. Penguin and box contributions
to $K \to\pi \nu\bar\nu$ in
  the LH model at $\ord (v^2/f^2)$ which are proportional to $v^4/f^4
  x_T x_L^2$.}\label{fig:class5}
\end{figure}

\begin{figure}[ht]
\vspace{0.10in}
\epsfysize=3.5in
\centerline{
\epsffile{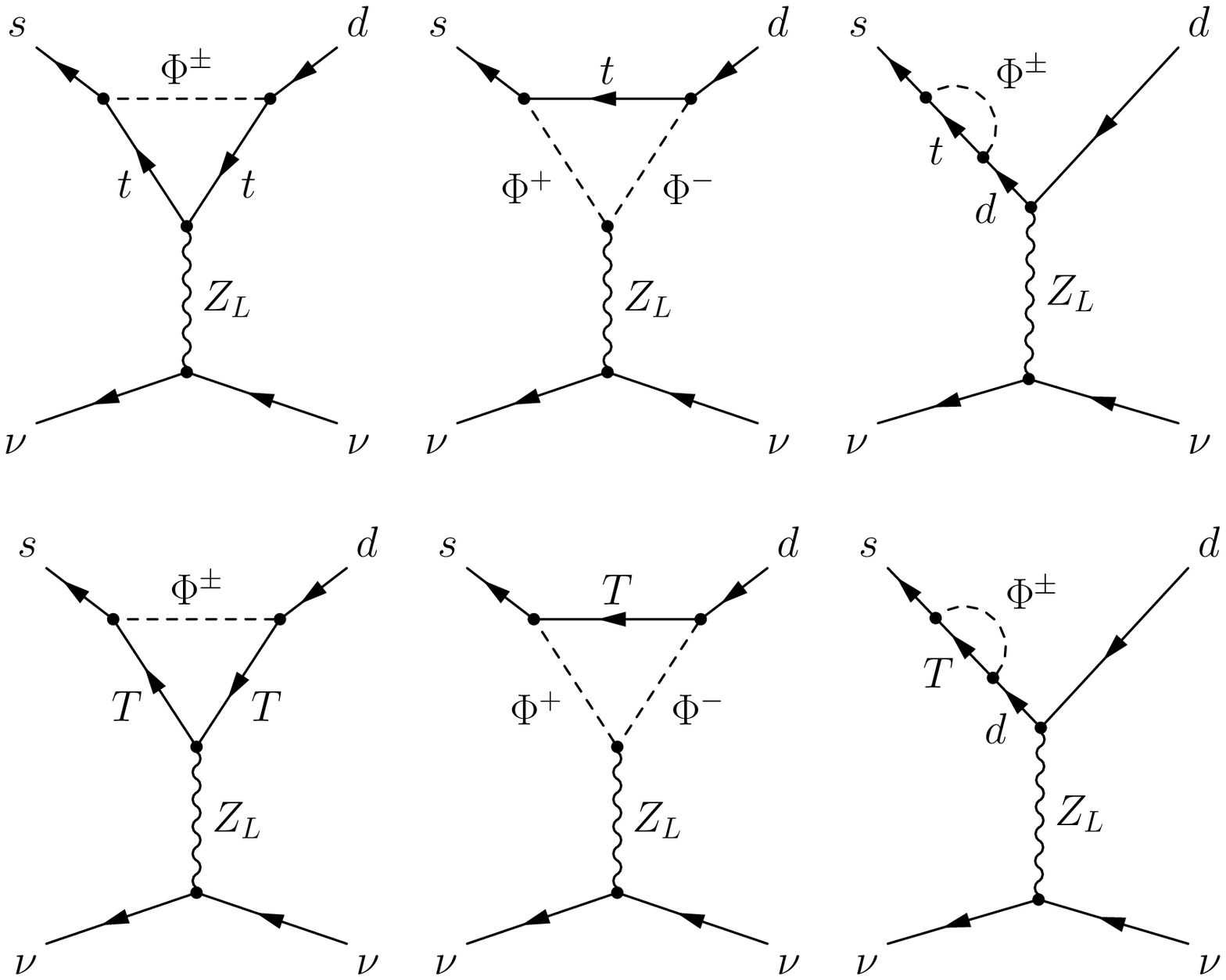}}
\vspace{0.01in}
\caption{Class 6. Penguin contributions
to $K \to\pi \nu\bar\nu$ in
  the LH model at $\ord (v^2/f^2)$ with internal charged scalars $\Phi^\pm$.
}\label{fig:class6}
\end{figure}

In the LH model the functions $X$ and $Y$ are modified through the
contributions of new penguin and box diagrams involving the heavy fields
$W_H$, $Z_H$, $A_H$, $T$ and $\Phi^\pm$. 
 In order to show transparently 
how the cancellation of most of the divergences takes place,
it is useful to group the diagrams contributing at $\ord(v^2/f^2)$ into 
six distinct classes which are shown in 
 Figs.~\ref{fig:class1}--\ref{fig:class6}.

 Class 1, displayed in  Fig.~\ref{fig:class1}, summarizes all diagrams with 
exclusively Standard Model particles. The circles
around the vertices of these diagrams indicate that the  $\ord(v^2/f^2)$
corrections to their vertices {\it without} the $x_L^2$ terms are 
considered. Using the leading order vertices one arrives at the
SM $X(x_t)$ function. Furthermore, the $W_LW_HZ_L$ triple vertex and
the ($W_L$,$A_H$) penguin diagrams with the standard top quark propagating 
belong to this class.

 Class 2 contains the contributions of the standard top quark in the 
($W_H$,$Z_L$) and ($W_L$,$Z_H$)
penguin diagrams, the
($W_L$,$W_H$) box diagram and the diagrams with the $W_L W_H Z_H$  and 
$W_H W_H Z_L$ triple vertices that are of order $v^2/f^2$. The whole
contribution of this class is proportional to the parameter $c^4$.

The penguin and box diagrams involving the heavy $T$
quark as well as the contributions of the standard top quark that
are proportional to $x_L^2$
are displayed in Fig.~\ref{fig:class3} and belong to Class 3.
The vertices of the standard top quark are marked by the diamonds
in this figure which
implies that here the terms in the vertices proportional to $x_L^2$,
excluded from Class 1, are considered. 
The approximate results for this class have been already presented in 
\cite{DEC05}. Here we present exact formulae at $\ord (v^2/f^2)$.

 The divergences of Class 3 involving the heavy top quark $T$ are
proportional to $v^2/f^2\, x_T/\varepsilon$. As the mass of the heavy $T$ is
of order $f/v$, these
singularities are of $\ord (1)$. This makes clear that
diagrams with singularities of the type $v^4/f^4\,x_T/\varepsilon$ also
have
to be considered and it turns out that the inclusion of these
divergences is essential for the removal of the singularities of the
whole $\ord (v^2/f^2)$ result except for the singularities discussed in 
Section~\ref{sec:Bardeen}. 
The relevant contributions of this type 
containing the
heavy top quark $T$ and being suppressed by $v^4/f^4$ are summarized in
Class 4 and Class 5.
Class 4 is very similar to Class 2 with the standard top quark
replaced by the heavy top quark and additional two diagrams with $t$
and $T$ exchanges. Class 5 contains diagrams of Class 1
with $t$ replaced by $T$ in the first five diagrams in Fig. 5 and with
corrections added to the last five diagrams in Fig.~\ref{fig:class3}
 as explicitly
indicated in Fig.~\ref{fig:class5}.

In the previous section we pointed out that due to the breakdown of the
custodial $SU(2)$ symmetry at $\ord (v^2/f^2)$ in the LH model, 
the SM relation (\ref{rel}) 
is replaced by (\ref{break}).
In the process of expressing $M_Z$ in the $Z$-penguin in terms of $M_W$
 all contributions of $\ord(1)$
belonging to  $Z_L$ vertices obtain $\ord (v^2/f^2)$
corrections. Explicitly, corrections to the contribution of the 
SM penguin diagrams of Class 1 and $Z_L$ penguins
with heavy top quark $T$ of Class 3 arise. We  find then
two additional contributions
\begin{align}\label{cust}
\Delta X_{\rm Custodial\, 1}&=\Delta Y_{\rm Custodial\, 1} = \frac{v^2}{f^2}
\frac{5}{4}(c'^2-s'^2)^2 C(x_t)_{\rm unitary},\\
\label{cust3}
\Delta X_{\rm Custodial\, 3}&=\Delta Y_{\rm Custodial\, 3} = \frac{v^4}{f^4}
\frac{5}{4}(c'^2-s'^2)^2 x_L^2 C(x_T)_{\rm Class\, 3},
\end{align}
with $C(x_t)_{\rm unitary}$ given in (\ref{Cunit}). For $ C(x_T)_{\rm
  Class\, 3}$ we obtain
\be
C(x_T)_{\rm
  Class\, 3}=-\frac{x_T}{16}\left(\frac{1}{\varepsilon}+\ln \frac{\mu^{2}}{M_{W_{L}}^{2}}\right)-\frac{3
  x_T}{32}+\frac{(-2+x_T) \log x_T}{16}.
\ee
It has to be emphasized that the inclusion of these two 
corrections resulting from the breakdown of custodial symmetry 
in the LH model is essential  for the removal of the $s'$ dependence as we will
show in the next section and removes some divergences.

Finally in Fig.~\ref{fig:class6}   we show the diagrams involving $\Phi^\pm$ that contribute
at $\ord(v^2/f^2)$.

\subsection{Analytic Results}

In order to explicitly show how the divergences cancel, we list in Appendix
\ref{sec:appB} the singularities in Class 1 to 3 in Table \ref{tab:class123},
and the ones of Class 4 and 5 in Table~\ref{tab:class45}.
The singularities in Class 6 are listed in Table~\ref{tab:class6}.
The entries of each class are arranged according to the position of
the corresponding diagrams in Figs.~\ref{fig:class1}--\ref{fig:class6}.
The variables $a$, $d_2$, $u$ and $a'$  are defined in (\ref{defa})
and (\ref{defu}).
 
The divergences in Class 2, Class 3 and Class 4 cancel separately
within each class. For the classes 1, 5 and 6 the situation is a bit
different. Some divergences of Class 1 and Class 5 can only be removed
in the sum of the singularities of both classes together with the inclusion
of the singularities due to the breakdown of the custodial $SU(2)$ symmetry,
\begin{align}
\label{custo1}
\frac{x_t}{\varepsilon}\frac{v^2}{f^2}
\left(-\frac{5}{64}(c'^{2}-s'^{2})^2\right),\\
\label{custo2}
\frac{x_T}{\varepsilon}\frac{v^4}{f^4}
x_L^2\left(-\frac{5}{64} (c'^{2}-s'^{2})^2\right),
\end{align}
which are also shown in Table \ref{tab:class123} and \ref{tab:class45},
respectively and can be obtained from (\ref{cust}) and (\ref{cust3}).
Further on, singularities of the standard top quark are canceled by those
of the heavy top quark with the use of relation (\ref{mT}) as
\be
x_L^2 \,\frac{x_T}{\varepsilon}=\frac{x_L}{1-x_L}\,\frac{x_t}{\varepsilon}\,
\frac{f^{2}}{v^{2}}.
\ee

However, as already stated at the beginning of our paper  singularities 
from classes 1 and 5 and the charged Higgs diagrams of Fig.~\ref{fig:class6}
 are left.
We find
\be\label{sing}
C_{\rm div}= \frac{x_t}{64}\frac{1}{1-x_L}\frac{v^2}{f^2}
\left(-\frac{S_{1}}{5}+S_{2}\right),
\ee
where
\begin{equation}\label{divergences}
S_{1} =  \frac{1}{\varepsilon} + \ln\frac{\mu^{2}}{M_{W_{L}}^{2}} \qquad \textrm{and} \qquad
S_{2} = \frac{1}{\varepsilon} + \ln\frac{\mu^{2}}{M_{\Phi}^{2}}.
\end{equation}
$S_{1}$ results from classes 1 and 5 and $S_{2}$
from charged Higgs diagrams. We will return to discuss these
singularities in the next Section. We caution
the reader that the logarithms associated with $1/\varepsilon$ have not been
explicitly shown in Tables \ref{tab:class123}-\ref{tab:class6}.

We can then write the results for $X$ and $Y$ in the LH model as
\begin{align}
X_{\rm LH}(x_t,z)&=X_{\rm SM}(x_t)+\Delta X_1+\Delta X_2+\Delta
X_3+\Delta X_4+\Delta X_5+\Delta X_6\\
Y_{\rm LH}(x_t,z)\,&=Y_{\rm SM}(x_t)+\Delta Y_1+\Delta Y_2+\Delta
Y_3+\Delta Y_4+\Delta Y_5+\Delta Y_6
\end{align}
where $z$ denotes collectively the parameters in the LH model to which we
will return below. The finite parts of the two corrections due to the custodial
relation given by (\ref{cust}) and (\ref{cust3}) were included into the 
$X$ and $Y$ functions of Class 1 and Class 5. We emphasize, that once these corrections are included the dependence on $s^{\prime}$ drops out.

For the six classes in question we find
\begin{align}
\Delta X_1&=\frac{v^2}{f^2}\,U_1,&
\Delta X_2&=c^4 \frac{v^2}{f^2}\,U_2\;\;\;=\frac{c^2}{s^2}\frac{1}{y} U_2,\\
\Delta X_3&=x_L^2\,\frac{v^2}{f^2}\,U_3,&
\Delta X_4&=x_L^2 c^4\frac{v^4}{f^4} U_4= x_L^2 \frac{c^2}{s^2} \frac{1}{y}
\frac{v^2}{f^2}\,U_4,\\
\Delta X_5&=x_{L}^{2} \frac{v^{4}}{f^{4}}\,U_5,&
\Delta X_6&=\frac{v^2}{f^2}\frac{x_t}{128}\frac{1}{1-x_L}
\left(1-2 x_L U_6(\hat x_T)\right)\\
\nonumber\\
\Delta Y_1&=\frac{v^2}{f^2}\,V_1,&
\Delta Y_2&=c^4\frac{v^2}{f^2}\,V_2\;\;\;=\frac{c^2}{s^2}\frac{1}{y} V_2,\\
\Delta Y_3&=x_L^2\,\frac{v^2}{f^2}\,V_3,&
\Delta Y_4&=x_L^2 c^4\frac{v^4}{f^4} V_4=x_L^2 \frac{c^2}{s^2} \frac{1}{y}
\frac{v^2}{f^2}\,V_4,\\
\Delta Y_5&=x_{L}^{2}\frac{v^4}{f^4}\,V_5 & \Delta Y_6&=\Delta X_6
\end{align}
with 

\begin{align}
U_1(x_t,y)\;\;=\;\;&-\frac{\left(1+4x_L\right)x_t}{320}S_{1} +\frac{\left(1+4x_L\right)\left(-7+x_t\right)x_t}{640\left(-1+x_t\right)}\nonumber\\
&+\frac{\left(1+4x_L\right)x_t\left(4-2x_t+x_t^2\right)\log x_t}{320\left(-1+x_t\right)^2}\nonumber\\
&-\frac{a x_t\left(11+4x_t\right)}{8\left(-1+x_t\right)}-\frac{3a x_t\left(-8+2x_t+x_t^2\right)\log x_t}{8\left(-1+x_t\right)^2}+\frac{3ax_t \log y}{8}\label{U1}\\
U_2(x_t,y)\;\;=\;\;&-\frac{x_t\left(4-7x_t\right)}{16(-1+x_t)}-\frac{3x_t\left(8-6x_t-x_t^2\right)\log x_t}{16(-1+x_t)^2}-\frac{x_t\log y}{4}\\
U_3(x_t,x_T)\;\;=\;\;&\frac{-3+2x_t-2x_t^2}{8(-1+x_t)}-\frac{x_t\left(-4-x_t+2x_t^2\right)\log x_t}{8(-1+x_t)^2}+\frac{\left(3+2x_t\right)\log x_T}{8}\\
U_4(x_T,y)\;\;=\;\;&\frac{3x_T y}{16\left(-x_T+y\right)}+\frac{3x_T y^2 \log x_T}{16(x_T-y)^2}-\frac{3x_T y^2 \log y}{16(x_T-y)^2}-\frac{x_T\log y}{16}\\
U_5(x_t,x_T)\;\;=\;\;&-\frac{\left(-3+4x_L\right)x_T}{320} S_{1}+\frac{\left(-7-12x_L+80x_L^2\right)x_T}{640}+\frac{\left(-3+4x_L\right)x_T\log x_T}{320}\nonumber\\
&+\frac{3 a x_T y\left(\log x_T - \log y\right)}{8\left(x_T-y\right)}\\
U_6(\hat x_T)\;\;=\;\;&-\frac{S_{2}}{x_{L}}+\frac{\hat x_T}{(1-\hat x_T)}+\frac{\hat x_T^2\log\hat x_T}{(1-\hat x_T)^2}\\
\nonumber\\
V_1(x_t,y)\;\;=\;\;&-\frac{\left(1+4x_L\right)x_t}{320}S_{1}+\frac{\left(1+4x_L\right)\left(-7+x_t\right)x_t}{640\left(-1+x_t\right)}\nonumber\\
&+\frac{\left(1+4x_L\right)x_t\left(4-2x_t+x_t^2\right)\log x_t}{320\left(-1+x_t\right)^2}\nonumber\\
&-\frac{a x_t\left(-13+4x_t\right)}{8\left(-1+x_t\right)}-\frac{3a x_t^2\left(2+x_t\right)\log x_t}{8\left(-1+x_t\right)^2}+\frac{3ax_t \log y}{8}\label{V1}\\
V_2(x_t,y)\;\;=\;\;&-\frac{x_t\left(4-7x_t\right)}{16(-1+x_t)}-\frac{3x_t^2\left(2-x_t\right)\log x_t}{16(-1+x_t)^2}-\frac{x_t\log y}{4}\\
V_3(x_t,x_T)\;\;=\;\;&\frac{\left(3+2x_t-2x_t^2\right)}{8(-1+x_t)}-\frac{x_t\left(2-x_t+2x_t^2\right)\log x_t}{8(-1+x_t)^2}+\frac{\left(3+2x_t\right)\log x_T}{8}\\
V_4(x_T, y)\;\;=\;\;&\frac{3x_T y}{16\left(-x_T+y\right)}+\frac{3x_T y^2 \log x_T}{16(x_T-y)^2}-\frac{3x_T y^2 \log y}{16(x_T-y)^2}-\frac{x_T \log y}{16}\\
V_5(x_t,x_T)\;\;=\;\;&-\frac{\left(-3+4x_L\right)x_T}{320}S_{1}+\frac{\left(-7-12x_L+80x_L^2\right)x_T}{640}+\frac{\left(-3 + 4x_L\right)x_T\log x_T}{320}\nonumber\\
&+\frac{3 a x_T y\left(\log x_T - \log y\right)}{8\left(x_T-y\right)},
\end{align}
where $S_{1}$ and $S_{2}$ have been defined in (\ref{divergences}) and
\be\label{xy}
x_{i} = \frac{m_{i}^{2}}{M_{W_L^\pm}^{2}},
\qquad y = \frac{M_{W_H^\pm}^{2}}{M_{W_L^\pm}^{2}},\qquad
\hat x_T= \frac{m_T^2}{M^2_{\Phi^\pm}}.
\ee
In our calculations, we considered all contributions to the order $v^2/f^2$.
This implies the neglection of higher order terms in the functions $U_i$
and $V_i$. As explained above, Class 4 and Class 5 even if suppressed by
$v^4/f^4$ factors, contribute to the order considered as they depend on
heavy masses and $x_T, y \propto f^2/v^2$.\\
The formulae for $\Delta X_i$ and $\Delta Y_i$ presented above are the 
main results of our paper.

\boldmath
\section{The Issue of Leftover Singularities}\label{sec:Bardeen}
\unboldmath

It may seem surprising that FCNC amplitudes considered
in the previous section contain residual ultraviolet
divergences reflected by the non-cancellation of the $1/ \varepsilon$
poles at $\mathcal{O}(v^{2}/f^{2})$ in our unitary gauge
calculation.
Indeed due to GIM mechanism the FCNC processes considered here vanish at 
tree level both in the SM and in the LH model in question. 
Therefore within the particle content of the low energy representation of 
the LH model there seems to be no freedom to cancel the left-over divergences 
as the necessary tree level counter terms are absent.

At first sight then one could worry that the remaining divergence is 
an artifact of the unitary gauge calculation. However, the fact that the 
dominant divergence comes from the gauge independent charged triplet Higgs
$\Phi^{\pm}$ contribution gives us a hint that the residual divergence is not 
an artifact of the unitary gauge but reflects the true sensitivity to the 
UV completion of the LH model and the presence of additional contributions to
the non-linear sigma model used as the effective field theory at low energy.

In order to put this hypothesis on a solid ground we have analyzed the 
divergent part of the amplitudes in the Feynman gauge. Then the box diagram 
contributions are finite and it is sufficient to concentrate on the penguin 
(vertex) contributions. In this context let us recall that in the SM the 
divergent contributions from penguin diagrams involving only quarks and 
gauge bosons are removed by the GIM mechanism as the divergent terms are mass 
independent. Some of the vertex diagrams with internal Goldstone bosons
are also divergent and being proportional to $m_{i}^{2}, (i = u, d, t)$ these 
divergences cannot be removed by GIM mechanism. Within the SM they cancel, 
however, due to gauge invariance and renormalizability of the theory.

In the LH model in Feynman gauge there are no divergences left from the pure 
gauge boson diagrams of classes 1-5 shown in 
Figs.~\ref{fig:class1}--\ref{fig:class5}. Note also that 
the divergence from the breakdown of the custodial symmetry is also absent 
as in the Feynman gauge, as seen in (\ref{C0xt}),
the SM function $C$ is finite. Thus the left-over 
divergences come only from the charged triplet Higgs contribution in 
Fig.~\ref{fig:class6} 
and two charged Goldstone bosons that now have to be included in the 
evaluation of the diagrams of classes 1-5. These are a charged vector 
Higgs boson which is responsible for the mass of $W_{H}$ and 
the usual charged doublet Higgs boson which gives mass to $W_{L}$. 
We confirm that the left-over divergence coming from these Goldstone 
boson contributions to classes 1-5 exactly reproduces the divergence
discovered in the corresponding unitary gauge calculation. Combined
with the charged triplet Higgs contribution we reproduce, in Feynman gauge, the
full divergence of (\ref{sing}).

To understand the meaning of these ultraviolet divergences it is 
important to recall that the LH model is a non-linear sigma model, 
an effective field theory that describes the low energy behavior of 
a symmetric theory below the scale where the symmetry is 
dynamically broken. In this region the currents associated with the 
dynamically broken generators are conserved by a cancellation between 
the quark charge form factor current and the Goldstone current. 
Quark currents will remain conserved even when the charge form factor is 
renormalized so long as the Yukawa coupling of the Goldstone bosons to 
the fermions has a corresponding renormalization. It is easy to confirm 
that this is exactly what happens in the non-linear sigma model used 
above to describe the Little Higgs theory and the divergence may be 
identified as a renormalization of the quark charges associated 
with neutral current processes. The subsequent gauging of the Little 
Higgs theory only rearranges the infrared structure of the theory but 
cannot modify the ultraviolet behavior. The divergence in the charge 
form factors is not a true ultraviolet divergence but reflects sensitivity 
to the UV completion of the theory.

This same mechanism can be observed in the phenomenological description of 
dynamical chiral symmetry breaking in QCD using a non-linear realization of 
the pseudo-scalar mesons as Goldstone bosons. Here the axial charges are 
dynamically broken but the axial vector currents remain conserved due to 
the Goldstone currents of pions. To apply this theory to the physical 
baryons, the axial charge of the baryon is observed to be renormalized, 
$G_{A} \sim 1.26 \neq 1$. This renormalization is consistent with a 
conserved axial vector current so long as the Goldstone coupling of the 
pions to the baryons is modified according to
the Goldberger-Treiman relation. In fact, the naive constituent quark model 
predicts an even larger value of $5/3$ for the axial charge of the baryon 
where the axial charge of the quark is taken to be $1$. As mentioned in 
the introduction, Peris \cite{Peris} has considered
the next-to-leading order chiral loop corrections to the axial charge form 
factors of the constituent quark. He uses a linear sigma model to regularize
the non-linear theory and finds a logarithmic sensitivity to the mass of 
the scalar partners to the pions reflecting the chiral splitting within 
the meson multiplet. In the non-linear version, his calculation would 
generate logarithmic divergences exactly analogous to the residual 
divergences we have found in the Littlest Higgs model. In this model the 
scalar partner masses cannot be larger than $4 \pi f$. Using this scale, 
Peris shows that the axial charge of the constituent quark is reduced by 
$20 \%$ in rough agreement with baryon phenomenology.

The value of the charge form factors of dynamically broken generators 
will depend on the ultraviolet completion of the Little Higgs model. 
The principal question concerns how the dynamical symmetry breaking is 
transmitted to the fermions. As a minimum, the symmetry breaking is reflected
through the Yukawa couplings of the Goldstone bosons to the fermions. 
In this case the next-to-leading corrections may be estimated from 
Goldstone loop corrections to the charge form factors and the scale of 
the logarithmic divergences should not be larger than $4 \pi f$. 
However, the light fermions may have a more complex relation to the 
fundamental fermions of the ultraviolet completion of the theory and 
the Little Higgs theory may have to include modifications of the charge form 
factors even at leading order, as in the case of the baryon 
where $G_{A} \neq 1$. We conclude that the residual logarithmic divergences 
found in Section 4 are a real physical effect, but they also indicate 
additional sensitivity to the ultraviolet completion of the Little Higgs 
models not usually included in the phenomenology of these models.

Assuming the minimal case discussed above, we estimate the contributions
of the logarithmically divergent terms to
the functions $X$ and $Y$. Removing $1/\varepsilon$ terms from (\ref{sing})
and setting $\mu = \Lambda$ we find

\begin{equation}
\Delta X_{div} = \Delta Y_{div} = \frac{x_{t}}{64} \frac{1}{1-x_{L}} \frac{v^{2}}{f^{2}} \left[\ln \frac{\Lambda^{2}}{M_{\Phi}^{2}} - \frac{1}{5} \ln \frac{\Lambda^{2}}{M_{W_{L}}^{2}}\right].
\end{equation}
Setting
\begin{equation}
\Lambda = 4 \pi f, \qquad m_{H} = 115\,\textrm{GeV}, \qquad v = 246\,\textrm{GeV}
\end{equation}
and using the values of $M_{W_{L}}$ and $m_{t}$ in Table \ref{tab:input} we find for $f/v = 5$ and $x_{L} = 0.8$

\begin{equation}
\Delta X_{div} = \Delta Y_{div} = 0.049,
\end{equation}
which should be compared with $X_{\textrm{SM}} \simeq 1.49$ and $Y_{\textrm{SM}} \simeq 0.95$. Thus for this choice of parameters the correction amounts to $3\%$ and $5\%$ for $X$ and $Y$, respectively. Larger values are obtained for $x_{L}$ closer to unity but such values are disfavoured by the measured value of $\Delta  M_{s}$ as discussed in the next section. Smaller values are found for larger $f$. In summary the effect of the logarithmic divergences turns out to be small. However, we would like to emphasize that our estimate takes only into account the contributions, where the fermions only couple to the Goldstone bosons through the mass terms, not the $G_{A}$-like terms, and the sensitivity to the ultraviolet completion of the LH model could in principle be larger than estimated here.

Few technical details on the issue of divergences  are given in Appendix \ref{Leftover}.

\boldmath
\section{Implications for Rare $K$ and $B$ Decays}\label{sec:Pheno}
\unboldmath

The most recent compendium of formulae for rare decays considered here, 
in terms of the functions $X$ and $Y$ can be found in two papers on rare 
decays in a model with one universal extra dimension \cite{BSW,BPSW}.
In order to obtain the relevant branching ratios in the LH model 
one only has to replace $X(x_t,1/R)$ and $Y(x_t,1/R)$ given there by
$X_{\rm LH}(v)$ and $Y_{\rm LH}(v)$ calculated here. Moreover, we included
the recently calculated NNLO QCD corrections \cite{BGHN06} and long
distance contributions \cite{Isidori05} to $\kpn$ that imply
$P_{c}=0.42 \pm 0.05$ for the charm contribution to this decay.

As we are mainly interested in the effects of the corrections coming from
LH contributions we will consider the ratios

\begin{eqnarray}
R_{+} \,\,\, &\equiv& \,\,\, \frac{Br\left(K^{+} \to \pi^{+} \nu \bar{\nu}\right)_{\textrm{LH}}}{Br\left(K^{+} \to \pi^{+} \nu \bar{\nu}\right)_{\textrm{SM}}},\label{ratiokppip}\\
R_{L} \,\,\, &\equiv& \,\,\, \frac{Br\left(K_{L} \to \pi^{0} \nu \bar{\nu}\right)_{\textrm{LH}}}{Br\left(K_{L} \to \pi^{0} \nu \bar{\nu}\right)_{\textrm{SM}}} =
\frac{Br\left(B \to X_{s,d} \nu \bar{\nu}\right)_{\textrm{LH}}}{Br\left(B \to X_{s,d} \nu \bar{\nu}\right)_{\textrm{SM}}} = \left[\frac{X_{\textrm{LH}}}{X_{\textrm{SM}}}\right]^{2},\label{ratioklpi0}\\
R_{s,d} \,\,\, &\equiv & \,\,\, \frac{Br\left(B_{s,d} \to \mu^{+} \mu^{-}\right)_{\textrm{LH}}}{Br\left(B_{s,d} \to \mu^{+} \mu^{-}\right)_{\textrm{SM}}} = \left[\frac{Y_{\textrm{LH}}}{Y_{\textrm{SM}}}\right]^{2},\label{ratioBmumu}
\end{eqnarray}
where with the values of $m_{t}$ and $M_{W_{L}}$ in Table \ref{tab:input} we
have

\begin{equation}
X_{\textrm{SM}} = 1.49, \qquad Y_{\textrm{SM}} = 0.95.
\end{equation}
In writing (\ref{ratioklpi0}) and (\ref{ratioBmumu}) we have assumed that
the values of the CKM parameters are the same in the SM and the LH model.
As both models belong to the class of MFV models for which the so-called
universal unitarity triangle exists \cite{MFV}, this assumption can certainly
be justified. Moreover, in principle CKM parameters can be determined from
tree level processes independently of new physics contributions. This
approach differs from the one followed in \cite{BPU04} where the CKM parameters
were determined using $B_{d}^{0}-\bar{B}_{d}^{0}$ mixing. As the relevant
one-loop function $S_{\textrm{LH}}$ in the LH model differs from the
$S_{\textrm{SM}}$, the CKM parameters turned out to be different in both
models in particular for $x_{L}$ close to unity. However, for $x_{L}$
close to unity $\left(\Delta M_{s}\right)_{\textrm{LH}}$ is significantly
larger than $\left(\Delta M_{s}\right)_{\textrm{SM}}$ in contradiction
with the recent CDF data that indicate $\Delta M_{s}$ to be smaller than
$\left(\Delta M_{s}\right)_{\textrm{SM}}$. The large non-perturbative
uncertainties in the evaluation of $\Delta M_{s}$ and also $\Delta M_{d}$
do not allow for a derivation of an upper bound on $x_{L}$ from
$B_{d,s}^{0}-\bar{B}_{d,s}^{0}$ mixings but clearly $x_{L}$ cannot be as
high as the $0.95$ used in \cite{BPU04, DEC05}. Therefore we will
choose $x_{L}\leq 0.8 $ in what follows.
Moreover, as stated above we will take the CKM parameters to be the same
for the SM and LH model and fixed to the central values collected in
Table \ref{tab:input}, where $\mtb=\mtb(m_t)$ in the
$\overline{\textrm{MS}}$ scheme. Then the ratios
in (\ref{ratioklpi0}) and (\ref{ratioBmumu}) only depend on the
one-loop functions $X$ and $Y$ and the dependence on the CKM parameters
is only present in (\ref{ratiokppip}) due to the relevant charm
contribution in $K^{+} \to \pi^{+} \nu \bar{\nu}$ in which the new
physics contributions are negligible.

\begin{table}[ht]
\renewcommand{\arraystretch}{1}\setlength{\arraycolsep}{1pt}
\center{
\begin{tabular}{|l|l|}
\hline 
{\small$\mtb= 163.8(32)\gev$} & {\small$|V_{ub}|=0.00423(35)$} \\
{\small$\mw= 80.425(38)\gev$} & {\small $\vcb = 0.0416(7)$\hfill\cite{BBpage}}\\\cline{2-2}
{\small$\alpha=1/127.9$} &{\small$\lambda=0.225(1)$\hfill\cite{CKM05}} \\\cline{2-2}
{\small$\sin^2 \theta_W=0.23120(15)$} & {\small $\gamma=71^{\circ}\pm16^{\circ}\hfill\cite{UTFIT}$}\\
\hline
\end{tabular}
}
\caption {{Values of the experimental and theoretical
    quantities used as input parameters.}}
\label{tab:input}
\renewcommand{\arraystretch}{1.0}
\end{table}

For the three new parameters $f$, $x_{L}$ and $s$ (see Section \ref{sec:LH}
for their definitions) we will choose the ranges
\be\label{parLH}
 {f}/{v} = 5 \,\, \textrm{or} \,\, 10, \quad 0.2 \leq x_{L} \leq 0.80, \quad 
0.3 \leq s \leq 0.95.
\ee
This parameter space is larger than the one allowed by other processes 
\cite{Logan}-\cite{PHEN6} which typically imply $f/v\ge 10$ or even higher. 
But we want to demonstrate that even for $f/v$ as low as $5$, the corrections
from LH contributions to $X$ and $Y$ are small.

\begin{figure}[ht]
\vspace{0.10in}
\epsfysize=8.0in
\centerline{
\epsffile{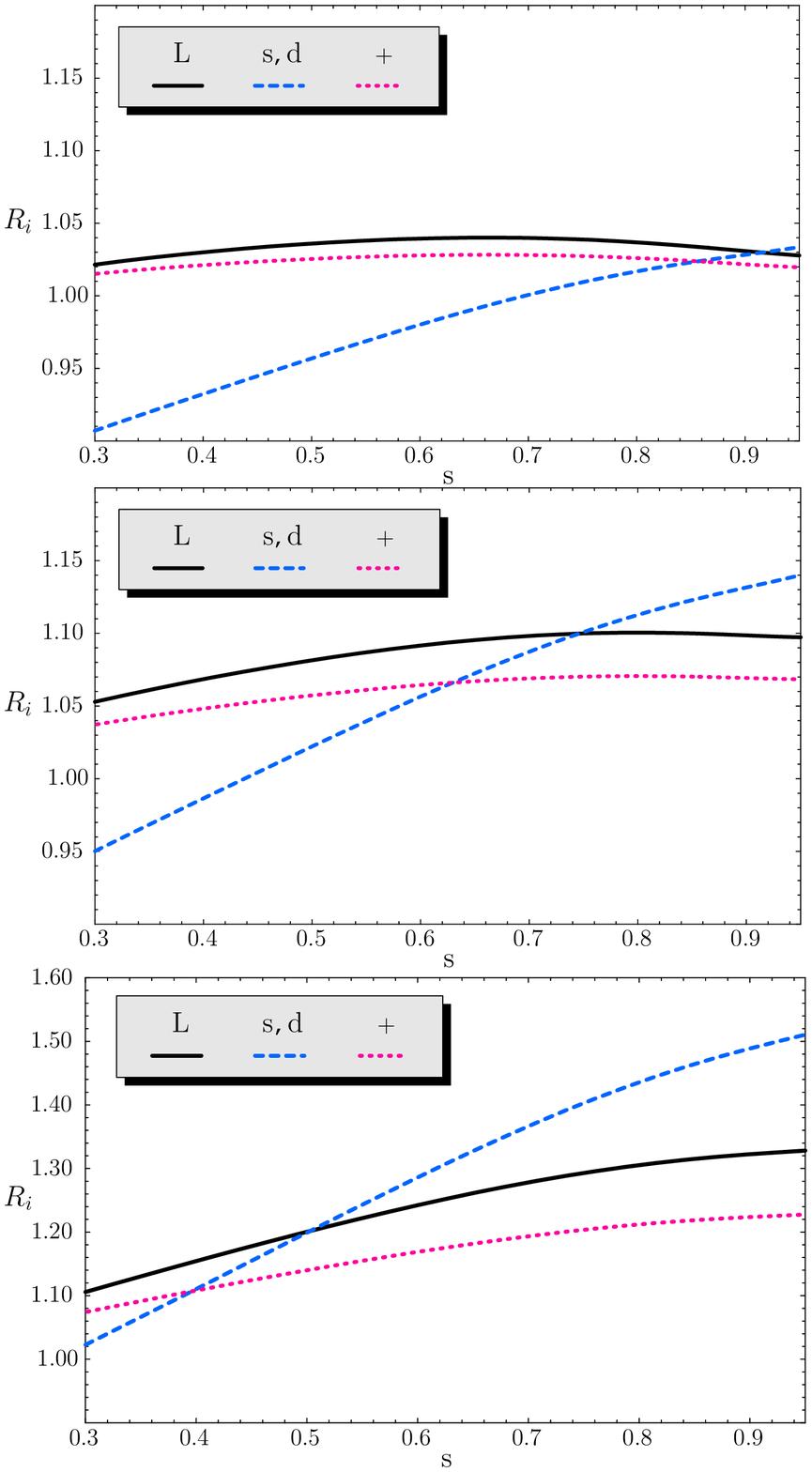}}
\vspace{0.01in}
\caption{Normalized branching ratios $R_{L}$, $R_{s,d}$, $R_{+}$ for
different $x_{L}=0.2, 0.5, 0.8$ (from top to bottom) and $f/v=5$.}\label{BR0.2}
\end{figure}

\begin{figure}[ht]
\vspace{0.10in}
\epsfysize=8.0in
\centerline{
\epsffile{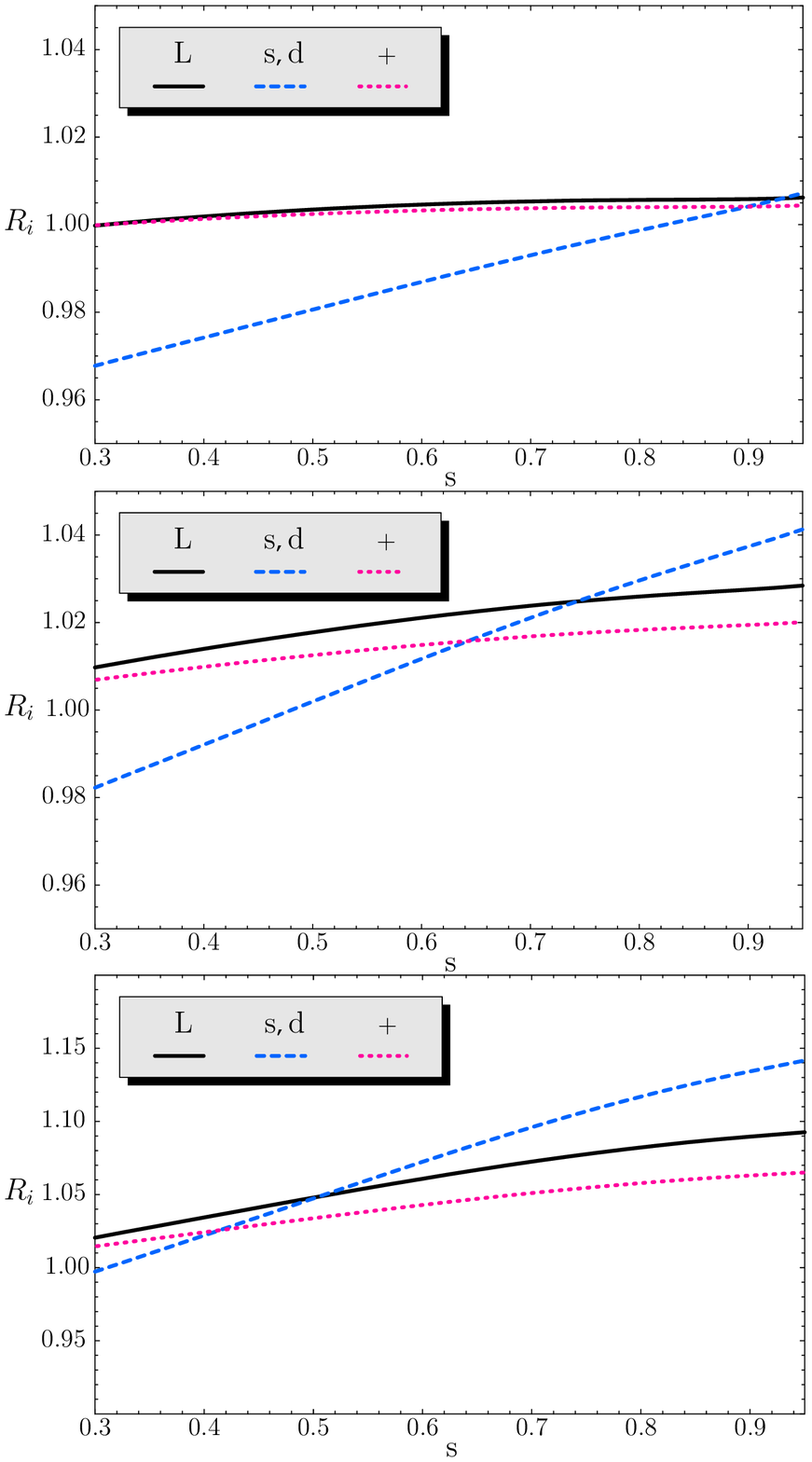}}
\vspace{0.01in}
\caption{Normalized branching ratios $R_{L}$, $R_{s,d}$, $R_{+}$ for
different $x_{L}=0.2, 0.5, 0.8$ (from top to bottom) and $f/v=10$.}\label{BR0.1}
\end{figure}

In Fig. \ref{BR0.2} we show the ratios (\ref{ratiokppip})-(\ref{ratioBmumu}) as
functions of $s$ for different values of $x_{L}$ and $f/v=5$. The
corresponding plots for $f/v=10$ are shown in Fig. \ref{BR0.1}.

We observe that $R_{+}$, $R_{L}$ and $R_{d,s}$ increase with increasing $s$ and
$x_{L}$. For $f/v=5$, $s=0.95$ and $x_{L}=0.8$ they reach $1.23$, $1.33$ and
$1.51$, respectively. However for $f/v=10$ they are all below $1.15$ and
consequently it will be difficult to distinguish the LH predictions for the
branching ratios in question from the SM ones.

\boldmath
\section{$B\to X_s\gamma$ Decay}\label{sec:Bsg}
\unboldmath
One of the most popular decays used to constrain new physics
contributions is the $B \rightarrow X_s \gamma$ decay for which
the measured branching ratio~\cite{BBpage}
\be Br(B \rightarrow
X_s \gamma)_\text{exp} = (3.52 \pm 0.30) \cdot 10^{-4}
\label{eq:bsgexp} \ee 
agrees well with the SM NLO
prediction~\cite{bsgSM1,bsgSM2}  
\be Br(B \rightarrow X_s
\gamma)_\text{SM} = (3.33 \pm 0.29) \cdot 10^{-4}\,,
\label{eq:bsgSM} \ee
both given for $E_\gamma>1.6\gev$ and the SM prediction for
$m_c(m_c)/m_b^{1S} = 0.26$. $Br(B\to X_d\gamma)$ is in the ballpark of
$1.5\cdot 10^{-5}$.

One should emphasize that within the SM this decay is governed by
the already well determined CKM element $|V_{ts}|$  so that
dominant uncertainties in~(\ref{eq:bsgSM}) result from the
truncation of the QCD perturbative series and the value of
$m_c(\mu)$ that enters the branching ratio first at the NLO level.
A very difficult NNLO calculation, presently in
progress~\cite{bsgSM2}, should reduce the error
in~(\ref{eq:bsgSM}) below $10$\%.

The effective Hamiltonian relevant for this decay within the SM is
given as follows
\be
\mathcal{H}_\text{eff}^\text{SM}(\bar b \rightarrow \bar s
\gamma) = - \frac{G_{F}}{\sqrt{2}} V_{ts} V_{tb}^* \left [ \sum_{i=1}^6 C_i(\mu_b)Q_i
+ C_{7\gamma}(\mu_b)Q_{7\gamma} + C_{8G}(\mu_b)Q_{8G}  \right ],
\label{eq:Heffbsg}
\ee

\newpage

\newpage

\newpage
where $Q_i$ are four-quark operators,
$Q_{7\gamma}$ is the magnetic photon penguin operator and $Q_{8G}$
the magnetic gluon penguin operator. The explicit expression for
the branching ratio $Br(B \rightarrow X_s \gamma)$ resulting
from~(\ref{eq:Heffbsg}) is very complicated and we will not present it
here. It can be found for instance in~\cite{bsgSM1}.

For our purposes it is sufficient to know that in the LO
approximation the Wilson coefficients $C_{7 \gamma}$ and $C_{8G}$ are given
at the renormalization scale $\mu_W=\mathcal{O}(M_W)$ as follows
\be
C_{7\gamma}^0(\mu_W) = - \dfrac{1}{2} D_0'(x_t)\,, \qquad C_{8G}^0(\mu_W) = -
\dfrac{1}{2} E_0'(x_t)\,,
\label{eq:C7g0C8G0}
\ee
with the explicit expressions for $D_0'(x_t)$ and $E_0'(x_t)$ given by

\begin{eqnarray}
D_{0}^{\prime}\left(x_{t}\right)\,\,& = &\,\,\frac{(8x_{t}^{3}+5x_{t}^{2}-7x_{t})}{12(x_{t}-1)^3}-\frac{(3x_{t}^{3}-2x_{t}^{2})\log x_{t}}{2(x_{t}-1)^4},\\
E_{0}^{\prime}\left(x_{t}\right)\,\,& = &\,\,\frac{(x_{t}^{3}-5x_{t}^{2}-2x_{t})}{4(x_{t}-1)^3}+\frac{3x_{t}^{2}\log x_{t}}{2(x_{t}-1)^4}.
\end{eqnarray}

In view of the importance of QCD corrections in this decay
we will include these corrections at NLO in the SM
part, but only at LO in the new contributions. This
amounts to including only corrections to the renormalization
of the operators in the LH part and eventually to increase
the scale $\mu_W$ to $\mu \approx 500 \gev$ at which the new particles are
integrated out. As the dominant QCD corrections to $Br(B \rightarrow X_s
\gamma)$ come anyway from the renormalization group evolution from $\mu_W$ down to $\mu_b=\mathcal{O}(m_b)$ and
the matrix elements of the operators $Q_2$ and $Q_{7\gamma}$ at $\mu_b$, these
dominant corrections are common to the SM and LH parts.

Within the LO approximation the new physics contributions to
$B \rightarrow X_s \gamma$ enter only through the modifications of the
functions $D_0'(x_t)$ and $E_0'(x_t)$.
This modification can be directly obtained by
changing the arguments in $D_0'(x_t)$ and $E_0'(x_t)$ and introducing 
corresponding factors that distinguish the LH model from SM contribution.
It is easy to see that the contributions with internal $(t,W^\pm_H)$ and 
$(T,W^\pm_H)$ are $\ord(v^4/f^4)$, while the contributions involving 
charged Higgs $\Phi^\pm$ enter first at even higher order. Consequently 
only the diagrams involving $W_L$, $t$ and $T$ contribute at
$\ord(v^2/f^2)$. We find then
\be
\left[D_0'(x_t,x_T)\right]_{\rm LH}=D_0'(x_t)+ \Delta D_0',
\qquad
 \left[E_0'(x_t,x_T)\right]_{\rm LH}=E_0'(x_t)+ \Delta E_0',
\ee
where
\be
\Delta D_0'
=\frac{v^2}{f^2}\left[x_L^2(D_0'(x_T)-D_0'(x_t))-2a\,D_0'(x_t)\right].
\ee
$\Delta E_0'$ is obtained from this equation by simply replacing 
$D_0'$ by $E_0'$.

The first calculation of $B \rightarrow X_s \gamma$ decay within the LH
model has been presented in \cite{BSG} and the result given above
confirms the one quoted in that paper. 

As in \cite{BSG} we find that the LH corrections amount to at most $3\%$
and are consequently smaller than the experimental and
theoretical uncertainties in (\ref{eq:bsgexp}) and (\ref{eq:bsgSM}).

\section{Conclusions}\label{sec:Summ}

In this paper we have presented for the first time a complete analysis of 
$\ord(v^2/f^2)$ contributions to rare $K$ and $B$ decays in the LH model of
\cite{LH1}-\cite{LH5}. The resulting corrections turned out to be small for values of 
the high energy scale $f=\ord (2-3)\tev$ as required by electroweak precision 
studies. While this is at first sight disappointing, one should recall the
upper bounds on rare decay branching ratios in MFV models \cite{Bobeth} that
do not allow for large departures from the SM predictions within the MFV
scenario. Thus the LH model considered here is consistent with these bounds.

On the technical side we have given a complete list of Feynman rules relevant
for the calculation of penguin and box diagrams that could be used for other 
processes. As a byproduct we have also presented for the first time the 
calculations of the $X$ and $Y$ functions in the unitary gauge both in the 
SM and the LH model. Some of the results obtained here can be used to 
calculate the T--even contributions to rare decays in the LH model with 
T--parity \cite{BBPTUW2}.

Probably the most interesting result of our paper is the left-over singularity
that signals some sensitivity of the final result to the UV completion of the
theory. A detailed discussion of this issue and of  possible implications 
of these findings for other LH models can be found in Section 5. A similar 
singularity has been found independently in the context of the study of 
electroweak precision constraints in \cite{Cheng}.

Large new physics effects have been found recently in the case of
$B^0_{d,s}-\bar B^0_{d,s}$ mixing in the LH model with T-parity,
where the presence of mirror fermions implies non-MFV interactions
\cite{Hubisz,BBPTUW}. The analysis of rare $K$ and $B$ decays presented
here will be generalized to this model in \cite{BBPTUW2}.

\vspace{0.3truecm}

\noindent
{\bf Acknowledgements}\\
\noindent
We would like to thank Monika Blanke, John Boersma, Bjoern Duling,
Ulrich Haisch, Heather Logan and Andreas Weiler for useful discussions. 
The work presented here was supported in part by the German 
Bundesministerium f\"ur
Bildung und Forschung under the contract 05HT4WOA/3 and by the German-Israeli
Foundation under the contract G-698-22.7/2002.
The research of William A. Bardeen is supported by Universities
Research Association, Inc. under contract No. DE-AC02-76CH03000
with the U.S. Department of Energy.




\newpage
\begin{appendix}
\setcounter{equation}{0}

\section{The Fermion Sector}
In order to  calculate $\ord(v^4/f^4 x_T)$ terms in classes 4 and 5 we
had to generalize the rules of \cite{Logan} by including certain higher 
order terms in $v/f$. Here we present some details of this derivation 
that also summarize the differences between our work and \cite{Logan} 
discussed in detail in \cite{BPU04}. 

The Yukawa Lagrangian for the top sector is given by \cite{Logan}
\begin{align}
\mathcal{L}_t=&\lambda_2 f \tilde t\tilde t'^c-i \lambda_1 t_3
\left[\sqrt{2}
  h^0+\frac{i}{f}(h^-\phi^++\sqrt{2}h^{0*}\phi^0)\right]u_3'^c\nonumber\\
&+i\lambda_1 \tilde t\left[-i f+\frac{i}{f}(h^+
  h^-+h^0h^{0*}+2\phi^{++}\phi^{--}+2 \phi^{+}\phi^{-}+2
  \phi^{0}\phi^{0*})\right]u_3'^c + h.c.\label{Lt}
\end{align}
where $t_3$ and $\tilde t$ are two components of the left handed
vector like fields $\chi_i=(b_3, t_3, \tilde t)$ replacing the third SM quark doublet and $u_3'^c$ and $\tilde
t'^c$ are the corresponding right handed singlets. $h$ and $\phi$
are the doublet and triplet scalar fields of the unbroken
$SU(2)_L\otimes U(1)_Y$. Spontaneous symmetry breaking via the vacuum
expectation values of the $h$ and $\phi$ fields, $\langle h^0\rangle =v/\sqrt{2}$ and
$\langle i \phi^0\rangle =v'$ generates the fermion masses. In the
following we set $v'=0$ as we did in our analysis.
Diagonalizing the Lagrangian (\ref{Lt}), one obtains the
left and right handed mass eigenstates of the light and the heavy top
quark. The field rotation, that has to be performed, is given by
\begin{align}
t_L&=c_L t_3-s_L \tilde t,&    t_R^c&=c_R u_3'^c-s_R \tilde t'^c,\\
T_L&=c_L t_3+s_L \tilde t,&  T_R^c&=c_R u_3'^c+s_R \tilde t'^c,
\end{align}
where we find
\begin{eqnarray}
s_R\,
&=&\,\frac{\lambda_1}{\sqrt{\lambda_1^2+\lambda_2^2}}\left(1-\frac{v^2}{f^2}\frac{\lambda_2^2}{\lambda_1^2+\lambda_2^2}\left(\frac{1}{2}-\frac{\lambda_1^2}{\lambda_1^2+\lambda_2^2}\right)+\ord(v^4/f^4)\right),\\
c_R\,&=&\,\frac{\lambda_2}{\sqrt{\lambda_1^2+\lambda_2^2}}\left(1+\frac{v^2}{f^2}\frac{\lambda_1^2}{\lambda_1^2+\lambda_2^2}\left(\frac{1}{2}-\frac{\lambda_1^2}{\lambda_1^2+\lambda_2^2}\right)+\ord(v^4/f^4)\right),\\
s_L\,&=&\,\frac{\lambda_1^2}{\lambda_1^2+\lambda_2^2}\frac{v}{f}\left(1+\frac{v^2}{f^2}\left(\!-\frac{5}{6}+\frac{1}{2}\frac{\lambda_1^4}{(\lambda_1^2+\lambda_2^2)^2}+2\frac{\lambda_1^2}{\lambda_1^2+\lambda_2^2}
    \left(1-\frac{\lambda_1^2}{\lambda_1^2+\lambda_2^2}\right)\right)\right)\nonumber\\
&&\,+\ord(v^4/f^4),\\
c_L\,&=&\,1-\frac{v^2}{f^2}\frac{1}{2}\frac{\lambda_1^4}{(\lambda_1^2+\lambda_2^2)^2}+\ord(v^4/f^4).
\end{eqnarray}
In order to obtain positive and real valued masses, 
it is necessary to absorb
a factor $-i$ into the $t_3$ field in (\ref{Lt}). A field
redefinition of this kind was suggested
but not performed by the authors of \cite{Logan}.
Then the masses of the light and the heavy top quark are given by
\begin{align}
m_t&=\frac{v \lambda_1
  \lambda_2}{\sqrt{\lambda_1^2+\lambda_2^2}}\left(1+\frac{v^2}{f^2}\left(-\frac{1}{3}+\frac{1}{2}\frac{\lambda_1^2}{\lambda_1^2+\lambda_2^2}\left(1-\frac{\lambda_1^2}{\lambda_1^2+\lambda_2^2}\right)\right)+\ord(v^4/f^4)\right),\label{mt1}\\
m_T&=f \sqrt{\lambda_1^2+\lambda_2^2}\left(1-\frac{v^2}{f^2}\frac{1}{2}\frac{\lambda_1^2}{\lambda_1^2+\lambda_2^2}\left(1-\frac{\lambda_1^2}{\lambda_1^2+\lambda_2^2}\right)+\ord(v^4/f^4)\right).\label{mt2}
\end{align}
The formulae of this section differ from the ones in \cite{Logan} in the
following points:
\begin{itemize}
\item
The masses $m_t$ and $m_T$ in (\ref{mt1}) and (\ref{mt2}) are real
valued and positive.
\item
$s_R$ and $c_R$ have the opposite sign in front of the last term of
the $\ord(v^2/f^2)$ expressions.
\end{itemize}

\newpage

\section{Tables of Singularities}\label{sec:appB}

\begin{table}[hbt]
\begin{center}
\caption[]{$1/\varepsilon$ Singularities to $\ord (v^2/f^2)$ of the
  Classes 1-3. The entries are arranged according to the position
of the corresponding diagrams in Fig.~\ref{fig:class1}-\ref{fig:class3}.}
\label{tab:class123}
\vspace{0.4truecm}
\begin{tabular}{|l|l|l|l|l|}
\hline
$\!\!\!\!\!\!$& Column 1& Column 2& Column 3 &Column 4\\
\hline
\multicolumn{5}{|l|}{Class 1.$\qquad$ Diagrams shown in Figure~\ref{fig:class1}. }\\
\hline

&\multicolumn{3}{l|}{}&\\
1&\multicolumn{3}{|l|}{$\frac{x_t}{\varepsilon}\left({\scriptstyle
      -}\frac{1}{16}{\scriptstyle +}\frac{v^2}{f^2}\left(\frac{5}{8}{\scriptstyle
        a}{\scriptstyle -}\frac{1}{8}{\scriptstyle u}{\scriptstyle
        +}\frac{5}{8}{\scriptstyle a'}{\scriptstyle +}\frac{1}{16}
   {\scriptstyle u x_L}
 \right)\right)$}&$\!\frac{x_t}{\varepsilon}\left(\!\frac{1}{16}{\scriptstyle +}\frac{v^2}{f^2}\left({\scriptstyle -}\frac{1}{4}{\scriptstyle a} \right)\right)\!\!\!$\\
&\multicolumn{3}{l|}{}&\\
\hline
&&\multicolumn{2}{|l|}{}&\\
2&$\!\frac{x_t}{\varepsilon}\frac{v^2}{f^2}\left({\scriptstyle
    -}\frac{3}{8}{\scriptstyle a}\right)$
&
\multicolumn{2}{|l|}{$\!\frac{x_t}{\varepsilon}\frac{v^2}{f^2}\!\left(\left({\scriptstyle
        c'^{2}}{\scriptstyle -}\frac{2}{5}\right)^2\
\!\!\!\left({\scriptstyle -}\frac{5}{16}\right){\scriptstyle -}\!\frac{1}{8}{\scriptstyle x_L}
  \!\left({\scriptstyle c'^2}{\scriptstyle -}\frac{2}{5}\right)\!\right)$}&
 \\
&&\multicolumn{2}{l|}{}&\\
\hline
&\multicolumn{4}{l|}{Custodial
  Correction:$\quad\frac{x_t}{\varepsilon}\frac{v^2}{f^2}\left(-\frac{5}{64}({\scriptstyle c'^2-s'^2})^{\scriptstyle 2}\right)$}\\
\hline
\hline
\multicolumn{5}{|l|}{Class 2.$\qquad$ Diagrams shown in
  Figure~\ref{fig:class2}.}\\
\hline
&&&&\\
1&$\frac{x_t}{\varepsilon}\frac{v^2}{f^2}\!\left({\scriptstyle
      -}\frac{1}{4}{\scriptstyle
      c^4}
  \right)\!\left({\scriptstyle 1-s_w^2}\right)\!\!$&$\frac{x_t}{\varepsilon}\frac{v^2}{f^2}\left(\frac{3}{8}{\scriptstyle
      c^4}
  \right)\left({\scriptstyle 1-s_w^2}\right)$&$\frac{x_t}{\varepsilon}\frac{v^2}{f^2}\!\left(\!{\scriptstyle
      -}\frac{3}{16}{\scriptstyle
      c^4}
  \right)\!({\scriptstyle 1-}\frac{2}{3}{\scriptstyle s_w^2})\!\!$&$\frac{x_t}{\varepsilon}\frac{v^2}{f^2}\left(\frac{1}{8}{\scriptstyle c^4}\right)$\\
&&&&\\
\hline
&&&&\\
2&$\frac{x_t}{\varepsilon}\frac{v^2}{f^2}\left({\scriptstyle
      -}\frac{1}{4}{\scriptstyle
      c^4}\right)$&$\frac{x_t}{\varepsilon}\frac{v^2}{f^2}\left({\scriptstyle
      -}\frac{3}{16}{\scriptstyle
      c^4}\right)$
&$\frac{x_t}{\varepsilon}\frac{v^2}{f^2}\left({\scriptstyle +}\frac{3}{8}{\scriptstyle
    c^4}\right)$ &
 \\
&&&&\\
\hline
\hline
\multicolumn{5}{|l|}{Class 3.$\qquad$ Diagrams shown in
  Figure~\ref{fig:class3}. }\\
\hline
&&&&\\
1& $\frac{x_t}{\varepsilon}\frac{v^2}{f^2}{\scriptstyle x_L^2}
\left(\frac{1}{2}{\scriptstyle -}\frac{1}{4}{\scriptstyle
    s_w^2}\right)$&$\frac{x_t}{\varepsilon}\frac{v^2}{f^2}{\scriptstyle x_L^2}\left({\scriptstyle -}\frac{3}{8}\right)({\scriptstyle 1-}{\scriptstyle
  s_w^2})$&$\frac{x_t}{\varepsilon}\frac{v^2}{f^2}{\scriptstyle x_L^2}
\left(\frac{3}{16}{\scriptstyle -}\frac{1}{8}{\scriptstyle s_w^2}\right)$\!\!&\\
&&&&\\
\hline
&&&&\\
2&$\frac{x_t}{\varepsilon}\frac{v^2}{f^2}{\scriptstyle x_L^2}
\left({\scriptstyle -}\frac{1}{16}\right)$&$\frac{x_T}{\varepsilon}\frac{v^2}{f^2}{\scriptstyle
   x_L^2}\left(\frac{1}{4}
    {\scriptstyle s_w^2}\right)$ &
  $\frac{x_T}{\varepsilon}\frac{v^2}{f^2}{\scriptstyle
    x_L^2}\left(\frac{3}{8}\right)({\scriptstyle 1-}
    {\scriptstyle s_w^2})\!\!$&\\
&&&&\\
\hline
&&&&\\
3&$\frac{x_T}{\varepsilon}\frac{v^2}{f^2}{\scriptstyle
  x_L^2}\left({\scriptstyle -}\frac{3}{16}{\scriptstyle +} \frac{1}{8}
   {\scriptstyle
     s_w^2}\right)\!\!\!$&$\!\left(\frac{x_t}{\varepsilon}{\scriptstyle +}\frac{x_T}{\varepsilon}\right)\frac{v^2}{f^2}{\scriptstyle x_L^2}\left({\scriptstyle -}\frac{1}{4}\right)$&$\frac{x_T}{\varepsilon}\!\frac{v^2}{f^2}{\scriptstyle x_L^2}\left(\frac{1}{16}\right)$&\\
&&&&\\
\hline
\end{tabular}
\end{center}
\end{table}

\newpage

\begin{table}[ht]
\caption[b]{$1/\varepsilon$ Singularities to $\ord (v^2/f^2)$ of the
  Classes 4 and 5.
The entries
  are arranged according to the position of the corresponding diagrams
  in Fig.~\ref{fig:class4} and \ref{fig:class5}.}
\label{tab:class45}
\vspace{0.4truecm}
\begin{center}
\begin{tabular}[b]{|l|l|l|l|l|}
\hline
$\!\!\!$$\!\!\!$& Column 1& Column 2& Column 3 &Column 4\\
\hline
\multicolumn{5}{|l|}{Class 4.$\qquad$ Diagrams shown in
  Figure~\ref{fig:class4}. }\\
\hline
&&&&\\
1& $\!\!\frac{x_T}{\varepsilon}
\frac{v^2}{f^2}\frac{c^2}{s^2}{\scriptstyle x_L^2}\frac{1}{y}\left(-\frac{1}{4}\right)$&$\!\!\frac{x_T}{\varepsilon}
\frac{v^2}{f^2}\frac{c^2}{s^2}{\scriptstyle x_L^2}\frac{1}{y}\left(\frac{3}{8}\right)$&$\!\!\frac{x_T}{\varepsilon}
\frac{v^2}{f^2}\frac{c^2}{s^2}{\scriptstyle x_L^2}\frac{1}{y}\left(-\frac{3}{16}\right)$&$\!\!\frac{x_T}{\varepsilon}
\frac{v^2}{f^2}\frac{c^2}{s^2}{\scriptstyle x_L^2}\frac{1}{y}\left(\frac{1}{8}\right)$\\
&&&&\\
\hline
&&&&\\
2& $\!\!\frac{x_T}{\varepsilon}
\frac{v^2}{f^2}\frac{c^2}{s^2}{\scriptstyle x_L^2}\frac{1}{y}\left(-\frac{1}{4}\right)$&$\!\!\frac{x_T}{\varepsilon}
\frac{v^2}{f^2}\frac{c^2}{s^2}{\scriptstyle x_L^2}\frac{1}{y}{\scriptstyle s_w^2 }\left(\frac{1}{4}\right)$&$\!\!\frac{x_T}{\varepsilon}
\frac{v^2}{f^2}\frac{c^2}{s^2}{\scriptstyle x_L^2}\frac{1}{y}{\scriptstyle (1-s_w^2)}\!\left(\frac{3}{8}\right)$&$\!\!\frac{x_T}{\varepsilon}
\!\frac{v^2}{f^2}\!\frac{c^2}{s^2}{\scriptstyle x_L^2}\!\frac{1}{y} {\scriptstyle \!(\frac{2}{3}s_w^2-1)}\!\!\left(\!\frac{3}{16}\!\right)\!\!\!$
\\
&&&&\\

\hline
\hline
\multicolumn{5}{|l|}{Class 5.$\qquad$ Diagrams shown in
  Figure~\ref{fig:class5}. }\\
\hline
&&&&\\
1&$\!\!\frac{v^2}{f^2}\frac{x_t}{\varepsilon}({\scriptstyle -}\frac{1}{10} {\scriptstyle
  x_L}{\scriptstyle +}\frac{1}{4}{\scriptstyle c'^2}
{\scriptstyle x_L})$&$\!\!\frac{v^4}{f^4}\frac{x_T}{\varepsilon}{\scriptstyle x_L^2}
(\frac{1}{20}{\scriptstyle -}\frac{5}{8}{\scriptstyle
  c'^2}{\scriptstyle +}$&$\!\!\frac{v^4}{f^4}\frac{x_T}{\varepsilon}{\scriptstyle
  x_L^2}({\scriptstyle -}\frac{1}{20}{\scriptstyle +}$&\\
&$\!\!{\scriptstyle +}\frac{v^4}{f^4}\frac{x_T}{\varepsilon}
{\scriptstyle x_L^2}(\frac{1}{2}{\scriptstyle
  c'^2}{\scriptstyle -}\frac{5}{4}{\scriptstyle
  c'^4})$&${\scriptstyle +}\frac{5}{4}{\scriptstyle
  c'^4}{\scriptstyle -}\frac{1}{20}{\scriptstyle x_L}{\scriptstyle +}\frac{1}{8} {\scriptstyle c'^2}
{\scriptstyle x_L})$&$\!\!\quad\quad\quad{\scriptstyle
  +}\frac{1}{4}\scriptstyle c'^2{\scriptstyle -}\frac{5}{16}
{\scriptstyle c'^4}\big)$&\\
&&&&\\
\hline
&&&&\\
2&$\!\!\frac{v^4}{f^4}\frac{x_T}{\varepsilon}{\scriptstyle
  x_L^2}({\scriptstyle -}\frac{3}{8}{\scriptstyle a})$&$\!\!\frac{v^4}{f^4}\frac{x_T}{\varepsilon}{\scriptstyle
  x_L^2}({\scriptstyle 1}{\scriptstyle -}{\scriptstyle s_w^2})$&$\!\!{\scriptstyle\!}\frac{v^4}{f^4}\!\frac{x_T}{\varepsilon}{\scriptstyle
 \! x_L^2}\!(\!\frac{1}{16}{\scriptstyle \!u}{\scriptstyle-}\!\frac{1}{16}{\scriptstyle
  \!u}{\scriptstyle x_L}\!{\scriptstyle-}\!\frac{5}{4}{\scriptstyle\! a'}\!{\scriptstyle-}\!\frac{1}{4}{\scriptstyle \!x_{\!L}^{\!2}}\!\!\!\!$&\\
&&$\!\!({\scriptstyle -}\frac{9}{8}{\scriptstyle a}{\scriptstyle +}\frac{3}{4} {\scriptstyle
  d_2}{\scriptstyle +}\frac{3}{8}{
\scriptstyle u}-\frac{15}{8}{\scriptstyle
a'})$&$\!\!{\scriptstyle +}{\scriptstyle
s_w^2}({\scriptstyle -}\frac{3}{4}{\scriptstyle a}{\scriptstyle +}\frac{1}{2}{\scriptstyle
d_2}{\scriptstyle +}\frac{1}{4}{\scriptstyle u}{\scriptstyle -}\frac{5}{4}{\scriptstyle a'}\!)\!)\!\!$&\\
&&&&\\
\hline
&&&&\\
3&$\!\!\frac{v^2}{f^2}\frac{x_t}{\varepsilon}{\scriptstyle
  x_L}({\scriptstyle -}\frac{1}{8}{\scriptstyle u}){\scriptstyle +}\frac{v^4}{f^4}\frac{x_T}{\varepsilon}{\scriptstyle
  x_L^2}$&$\!\!\frac{v^4}{f^4}\!\frac{x_T}{\varepsilon}{\scriptstyle
  \!x_L^2}(\!\frac{3}{4}{\scriptstyle \!a}{\scriptstyle
  -}\frac{3}{8}{\scriptstyle \!d_2-}\frac{1}{4}{\scriptstyle
  \!u+}\frac{5}{4}{\scriptstyle
  \!a'\!+}\!\!$&$\!\!\frac{v^4}{f^4}\frac{x_T}{\varepsilon}{\scriptstyle
  x_L^2}({\scriptstyle -}\frac{1}{4}{\scriptstyle
a+}\frac{1}{8}{\scriptstyle d_2})$&\\
&$({\scriptstyle a}{\scriptstyle -}\frac{1}{2}{\scriptstyle
  d_2}{\scriptstyle -}\frac{1}{4}{\scriptstyle u}{\scriptstyle +}\frac{1}{4}{\scriptstyle
  x_L^2}{\scriptstyle +}\frac{5}{2}{\scriptstyle a'})\!\!$&${\scriptstyle
  s_w^2} ({\scriptstyle -}\frac{3}{8}{\scriptstyle
    a+}\frac{1}{4}{\scriptstyle d_2+}\frac{1}{8}{\scriptstyle
    u-}\frac{5}{8}{\scriptstyle a'}))\!\!$&&\\
&&&&\\
\hline
&\multicolumn{4}{l|}{Custodial
  Correction:$\quad\frac{x_T}{\varepsilon}\frac{v^4}{f^4}{\scriptstyle
    x_L^2}\left(-\frac{5}{64}({\scriptstyle c'^2-s'^2})^{\scriptstyle 2}\right)$}\\
\hline
\end{tabular}
\end{center}
\end{table}

\begin{table}[ht]
\caption[b]{$1/\varepsilon$ Singularities to $\ord (v^2/f^2)$ in 
  Class 6
The entries
  are arranged according to the position of the corresponding diagrams
  in Fig.~\ref{fig:class6}.}
\label{tab:class6}
\vspace{0.4truecm}
\begin{center}
\begin{tabular}[b]{|l|l|l|l|}
\hline
\multicolumn{4}{|l|}{Class 6.$\qquad$ Diagrams shown in
  Figure~\ref{fig:class6}. }\\
\hline
$\!\!$Line$\!\!$& Column 1& Column 2& Column 3 \\
\hline
&&&\\
1&$\frac{x_t}{\varepsilon}\frac{v^2}{f^2}\!\left({\scriptstyle
      -}\frac{1}{48}{\scriptstyle
      s_w^2}\right)\!\!$
   &$\frac{x_t}{\varepsilon}\frac{v^2}{f^2}\left(\frac{1}{32}{\scriptstyle
      s_w^2}\right)$&
$\frac{x_t}{\varepsilon}\frac{v^2}{f^2}\!\left(\!{\scriptstyle
      }\frac{1}{64} \right)
\!({\scriptstyle 1-}\frac{2}{3}{\scriptstyle s_w^2})
\!\!$\\
&&&\\
\hline
&&&\\
2&$\frac{x_t}{\varepsilon}\frac{v^2}{f^2}\!\left({\scriptstyle
      -}\frac{1}{48}{\scriptstyle
      s_w^2}\right)\frac{x_L}{1-x_L}\!\!$
   &$\frac{x_t}{\varepsilon}\frac{v^2}{f^2}\left(\frac{1}{32}{\scriptstyle
      s_w^2}\right) \frac{x_L}{1-x_L}$&
$\frac{x_t}{\varepsilon}\frac{v^2}{f^2}\!\left(\!{\scriptstyle
      }\frac{1}{64} \right)
\!({\scriptstyle 1-}\frac{2}{3}{\scriptstyle s_w^2})\frac{x_L}{1-x_L}
\!\!$\\
&&&\\
\hline
\end{tabular}
\end{center}
\end{table}

\newpage

\section{Comments on the Leftover Singularities}\label{Leftover}

\setcounter{equation}{0}

In Section \ref{sec:Bardeen} we found that in the amplitudes for FCNCs leftover singularities remained. As pointed out these divergent terms do not depend on the choice of a special gauge. This led to the conclusion that these divergences are a real physical effect and can be identified as a renormalization of the quark charges. In our calculation the divergent quark vertex contribution, using fundamental gauge fields, reads

\begin{equation}
V_{quark} = \frac{1}{4} \left(\lambda_{1} v\right)^{2} \frac{1}{\left(4 \pi\right)^2} \frac{1}{\varepsilon} \tilde{V}_{quark} \gamma_{\mu}\left(1-\gamma_{5}\right), \label{divergentresult}
\end{equation}
where $\tilde{V}_{quark}$ is given by
\begin{equation}
\tilde{V}_{quark}=\frac{4}{v^{2}}\left\{\frac{1}{4} \left(g_{1} W_{1}^{3}\right) - \frac{1}{4} \left(g_{2} W_{2}^{3}\right) + \frac{1}{20} \left(g_{1}^{\prime} B_{1}\right) - \frac{1}{20} \left(g_{2}^{\prime} B_{2}\right)\right\}. \label{weakeigenstates}
\end{equation}
Rewriting this result in a mass diagonal basis then yields

\begin{equation}
\tilde{V}_{quark} = \frac{1}{v^{2}} \frac{g}{s c} Z_{H} + \frac{1}{5}\frac{1}{v^{2}} \frac{g^{\prime}}{s^{\prime} c^{\prime}} A_{H} + \frac{1}{2} \frac{1}{f^{2}} \left[-c^{2} + s^{2} + c^{\prime 2} - s^{\prime 2}\right] \sqrt{g^{2} + g^{\prime 2}} Z_{L} \label{masseigenstates}.
\end{equation}
As it can be seen from (\ref{masseigenstates}) the coefficient of the physical $Z$-boson is suppressed in two different scenarios:
\begin{itemize}
\item if the gauge couplings of the two gauge groups are identical, i.e. $c = s$ and $c^{\prime} = s^{\prime}$. This, for example, happens in the case of the T-even sector of the Littlest Higgs Model with T-parity \cite{BBPTUW2}, where $c$ and $s$ are set to $c = s = 1/\sqrt{2} = c^{\prime} = s^{\prime}$.

\item if one of the product gauge groups is strongly coupled, i.e. if $c$, $c^{\prime}$ $\approx$ 1 or $s$, $s^{\prime}$ $\approx $ 1.
\end{itemize}

\end{appendix}




%
%
%
\end{document}